\documentclass[
preprintnumbers,
preprint,
a4paper,
showpacs, amsmath, amssymb, nofootinbib
]{revtex4}
\usepackage{amsmath,amssymb}
\usepackage{graphicx}
\begin{document}
\title{ Improved analysis of black hole formation\\ in high-energy particle collisions}
\author{Hirotaka Yoshino}
\email{yoshino@gravity.phys.nagoya-u.ac.jp}
\affiliation{Department of Physics, Graduate School of Science, Nagoya
University, Chikusa, Nagoya 464-8602, Japan}
\author{Vyacheslav S. Rychkov}
\email{rychkov@science.uva.nl}
\affiliation{Insituut voor Theoretische Fysica, Universiteit van Amsterdam,
  Valckenierstraat 65, 1018XE Amsterdam, The Netherlands}
\preprint{hep-th/0503171}
\preprint{DPNU-05-01}
\preprint{ITFA-2005-09}
\date{May 2, 2005}
\begin{abstract}
 We investigate formation of an apparent horizon (AH)
 in high-energy particle collisions in four- and higher-dimensional
 general relativity, motivated by TeV-scale gravity scenarios.
 The goal is to estimate the prefactor in the geometric
 cross section formula for the black hole production.
 We numerically construct AHs on the future light cone
 of the collision plane. Since this slice lies to
 the future of the slice used previously by Eardley and Giddings (gr-qc/0201034)
 and by one of us and Nambu (gr-qc/0209003), we are able
 to improve the prefactor estimates. The black hole
 production cross section increases by 40-70\%
 in the higher-dimensional cases, indicating larger
 black hole production rates
 in future-planned accelerators than previously estimated.
 We also determine the mass and the angular momentum
 of the final black hole state, as allowed by the area theorem.
\end{abstract}
\maketitle


\section{Introduction}

Several scenarios in which the fundamental Planck energy could be
$O({\rm TeV})$ have been proposed. In these scenarios, our space
is a 3-brane in large~\cite{ADD98} or warped~\cite{RS99} extra
dimension(s), and gauge particles and interactions are confined on
it. If this is the case, a black hole smaller than the
extra-dimension size is well described as a $D$-dimensional black
hole centered on the brane (where $D$ is the total number of the
large dimensions
\footnote{Thus $D=4+n$, where $n$ is the total number of large {\it
extra} dimensions.}), and its gravitational radius is far larger than
that of a usual black hole with the same mass. This implies that
such black holes could be produced using future-planned
accelerators, because the gravitational interaction becomes
dominant in particle collisions above the TeV scale, and the black
hole production cross section
\begin{equation}
\label{rough} \sigma_{\rm BH}\sim \pi \left[r_h(2\mu)\right]^2
\end{equation}
becomes sufficiently large. Here, $\mu$ is the energy of
each incoming particle in the center-of-mass frame of the collision
and $r_h(2\mu)$ is the gravitational radius
of a $D$-dimensional Schwarzschild black hole of mass $2\mu$,
given by \cite{MP86}
\begin{equation}
\label{rh}
r_h(2\mu)=\left[\frac{16\pi
G_D(2\mu)}{(D-2)\Omega_{D-2}}\right]^{1/(D-3)},
\end{equation}
where $G_D$ is the $D$-dimensional gravitational constant, and
$\Omega_{D-2}$ is the $(D-2)$-area of a unit sphere.

The phenomenology of black hole production in accelerators was
first discussed in~\cite{BHUA} (for reviews, see~\cite{reviews};
for a related issue of black hole production in cosmic rays, see
e.g.~\cite{cosmic}). There are four stages in the time evolution
of a produced black hole. The first one is horizon formation in
the particle collision. Next the balding phase follows, in which
classical emission of gravitational waves occurs, and the produced
black hole relaxes to a $D$-dimensional Kerr black hole, whose
metric was found by Myers and Perry~\cite{MP86}. The third stage
is the evaporation phase, in which the black hole evaporates due
to the Hawking radiation and superradiance\footnote{See
\cite{Stojkovic} and references therein for an interesting recent
discussion of the role of superradiance.}. The particles emitted
in this process are observed as the signals in accelerators. As
the black hole evaporates, its mass approaches the Planck mass. In
this Planck phase, quantum gravity effects will become important.

The Planck phase may lead to a number of unexpected phenomena as
predicted by string theory, non-commutative geometry,
etc.~\cite{quantum}. On the other hand, the first three phases are
well described by classical or semi-classical gravity. Quantitative
predictions concerning these phases are
important, since such predictions are necessary to test
the validity of higher-dimensional general relativity.
Furthermore, since quantum gravity effects will be observed by the
difference from the semi-classical signals, precise prediction of
the semi-classical signals is required.

Related to the evaporation phase, several studies of the greybody
factors of $D$-dimensional black holes are
available~\cite{greybody} (see also \cite{related} for related issues).
On the other hand, it is also necessary
to investigate the process of the black hole production and its
relaxation, because the black hole production cross section is
directly related to the black hole production rate in
accelerators. Furthermore, the classical gravitational radiation
determines the mass and angular momentum of the final state of the
produced black hole, which provides the initial conditions for the
evaporation process. Hence, the study of the high-energy
two-particle system is an important problem.

\begin{figure}[bt]
\centering
{
\includegraphics[width=0.25\textwidth]{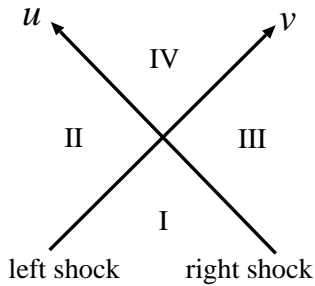}
}
\caption{Schematic picture of the spacetime of colliding high-energy particles. }
\end{figure}

The high-energy two-particle system in four dimensions has been
discussed to some extent before the appearance of the brane world
scenarios. The metric of one high-energy particle was obtained by
Aichelburg and Sexl~\cite{AS71} by boosting the Schwarzschild
black hole to the speed of light with fixed energy $\mu$. The
gravitational wave emission in the axisymmetric system of two
combined Aichelburg-Sexl particles was studied by
D'Eath~\cite{D'Eath} and D'Eath and Payne~\cite{DP92} (summarized
in \cite{D96}). A schematic picture of the spacetime with two
Aichelburg-Sexl particles is shown in Fig. 1. Two particles
collide at the speed of light. The gravitational field of each
incoming particle is infinitely Lorentz-contracted and forms a
shock wave. Except at the shock waves, the spacetime is flat
before the collision (i.e., regions I, II, and III). After the
collision, the two shocks nonlinearly interact with each other,
and the spacetime within the future lightcone of the collision
(i.e., region IV) becomes highly curved. The ultimate goal would
be to clarify the structure of region IV. If this is possible, the
black hole production cross section and the gravitational waves
emitted in the relaxation process could be determined. But this
analysis is difficult because of the quite complicated structure
of the gravitational field, and no one has succeeded in deriving
the metric in region IV even numerically.

Nonetheless, one can estimate the lower bound on the black hole
production cross section only with the knowledge of regions I, II,
and III. This can be done by finding an apparent horizon (AH),
because the AH existence is a sufficient condition for the event
horizon (EH) formation \cite{W84} (assuming the cosmic
censorship~\cite{P69}). As mentioned in \cite{D'Eath,DP92,D96}
(see also \cite{EG02}), Penrose (1974, unpublished) constructed an
AH on the slice $u=0,v<0$ and $v=0,u<0$ in the head-on collision
case in four-dimensional spacetime. Because this AH has intrinsic
geometry of combined two flat disks, it is often called the
Penrose flat disks. Eardley and Giddings~\cite{EG02} extended the
AH solution of Penrose flat disks to positive impact parameters.
They analytically derived the maximal impact parameter
$\hat{b}_{max}$ for the AH formation in a grazing collision in the
four-dimensional spacetime. Subsequently, one of us and
Nambu~\cite{YN03} extended this analysis to higher-dimensional
spacetimes. The values of $\hat{b}_{max}$ in $D$-dimensional cases
were obtained numerically, and they can be well approximated by
$\hat{b}_{max}\simeq 1.5\times 2^{-1/(D-3)}r_h(2\mu)$.

The AH method provides a lower bound on the true collision cross
section. This lower bound depends on the slice used to determine the AH
and becomes larger if a future slice is chosen. Indeed, the maximal
impact parameter of the AH formation will be larger for such a slice
(simply because it is possible that, for a given impact parameter, an AH
has not yet formed on the old slice, while it forms by the time a later
slice is reached).
The lower bound would asymptotically approach the exact cross section as
we move into the far future.
Because of the monotonic growth, the further we move into the future,
the smaller the difference between the true cross section and the
estimate provided by the AH method would become.

After the works of \cite{EG02, YN03}, one of us raised doubts in
the validity of the setup of the high-energy two-particle
system~\cite{R04}, because of possible strong curvature effects in
colliding shocks. However, this problem was later shown to be an
artifact of the unphysical classical point-particle limit: for a
particle described by a small quantum wavepacket large curvatures
do not arise~\cite{GR04} (see also \cite{R04-2}). Roughly, if a
wave packet of Planck size $\delta z$ is taken, curvature remains
sufficiently small, while corrections to the Aichelburg-Sexl
geometry are $O(\delta z/r_h)\ll 1$. This argument justified the
use of the Aichelburg-Sexl two-particle system to compute the
black hole production cross section in elementary particle
collisions. See also \cite{Y05}, \cite{R04-3} for other
characteristics of incoming particles that could affect black hole
formation.

In light of the above discussion, the purpose of this paper is as
follows. In the previous analysis~\cite{EG02, YN03}, the AH was
constructed on the union of the two incoming shocks $u=0, v<0$ and
$v=0, u<0$ (referred below as the old slice). However, it is clear
from Fig.~1 that this slice is not at all optimal in the sense
that there exist other slices within regions I, II, III, located
to the future of the old slice.  Motivated by
this observation, we proceed with the AH analysis on the slice of
the future light cone of the shock collision plane, given by the
union of the outgoing shocks $u=0,v>0$ and $v=0, u>0$ (referred
below as the new slice). This slice is optimal in the sense that
it is the future-most slice that can be taken without the
knowledge of region IV. By this analysis, we will improve the
lower bound on the cross section of the black hole production. In
addition, using the area theorem~\cite{H71}, we will find
restrictions on the mass $M$ and the angular momentum $J$ of the
final state (i.e. the produced black hole after the balding
phase). This part of the analysis is new compared to \cite{YN03},
and provides indirect information about the spacetime structure of
region IV of the Aichelburg-Sexl two-particle system.

This paper is organized as follows. In the next section, we
explain the system setup and derive the AH equation and the
boundary conditions in the new slice. Then we present the analytic
solution of the AH equation in the head-on collision case and
explain the numerical method for the more physically important
grazing collision case. In Sec. III, we present our numerical
results. We summarize the results for the maximal impact parameter
and discuss the mass $M$ and angular momentum $J$ of the final
state, as allowed by the area theorem. Sec. IV is devoted to the
summary and discussion.

Our new lower bounds on $\sigma_{\rm BH}$, most precise to date,
are summarized in Table II.

\section{Apparent horizons in the high-energy particle system}

\subsection{System setup}

We begin by reviewing the Aichelburg-Sexl metric, describing the
gravitational field of a high-energy particle. Following the
analysis in~\cite{EG02, YN03}, we use the metric of a massless
point particle of \cite{AS71, D'Eath, DH, EG02} that is obtained
by boosting the Schwarzschild black hole to the speed of light
with fixed energy $\mu=\gamma M$. The result is
\begin{equation}
ds^2=-d\bar{u}d\bar{v}+d\bar{r}^2+\bar{r}^2 d\bar{\Omega}_{D-3}^2
+\Phi(\bar{r})\delta(\bar{u})d\bar{u}^2,
\label{discontinuous}
\end{equation}
\begin{equation}
\Phi(\bar{r})=\left\{
\begin{array}{ll}
-2\log\bar{r} & (D=4),\\
{2}/{(D-4)\bar{r}^{D-4}} & (D\geq 5).
\end{array}
\right.
\end{equation}
Here we adopt $r_0=\left(8\pi
G_D\mu/\Omega_{D-3}\right)^{1/(D-3)}$ as the unit of length, which
is close to $r_h(2\mu)$. The delta function in
Eq.~\eqref{discontinuous} indicates that the coordinate system is
discontinuous at $\bar{u}=0$, and that a distributional Riemann
curvature (i.e., a gravitational shock wave) is located there. The
continuous coordinates are introduced by
\begin{align}
\bar{u}&=u,\\
\bar{v}&=
\left\{
\begin{array}{ll}
v-2\log r\theta(u)+{u\theta(u)}/{r^2}& (D=4),\\
v+{2\theta(u)}/{(D-4)r^{D-4}}+{u\theta(u)}/{r^{2D-6}} & (D\ge 5),
\end{array}
\right.\\
\bar{r}&=r\left(1-\frac{u}{r^{D-2}}\theta(u)\right),\\
\bar{\phi}_i&=\phi_i,
\end{align}
where $\bar{\phi}_i$ are the coordinates on the $(D-3)$-sphere.
The metric becomes \cite{D'Eath,R04}
\begin{equation}
ds^2=-dudv
+\left[1+(D-3)\frac{u}{r^{D-2}}\theta(u)\right]^2dr^2
+r^2\left[1-\frac{u}{r^{D-2}}\theta(u)\right]^2d\Omega^2_{D-3},
\end{equation}
where $\theta(u)$ denotes the Heaviside step function. Note that
$u=r^{D-2}$ is a coordinate singularity for $u\ge 0$, because the
$(D-3)$-sphere shrinks to zero size. Thus the $\bar{u}>0$ region
is mapped onto the $0<u\le r^{D-2}$ region by this coordinate
transformation.

\begin{figure}[tb]
\centering
{
\includegraphics[width=0.5\textwidth]{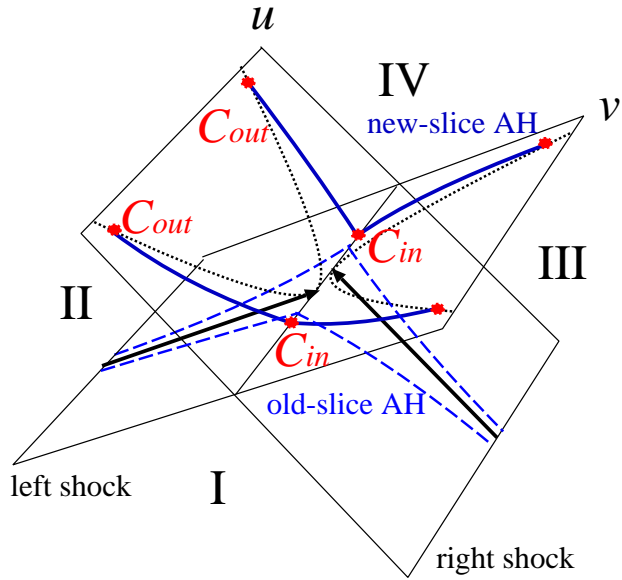}
} \caption{Schematic picture of the spacetime of colliding
high-energy particles with $(D-3)$ dimensions suppressed. The
schematic shape of AH on the new slice ($u>0,v=0$ and $v>0,u=0$)
is shown by solid lines, while the AH on the old slice ($u<0,v=0$
and $v<0,u=0$) is shown by dashed lines. Dotted lines indicate
coordinate singularities.}
\end{figure}

By causality, we can construct the metric of a high-energy
two-particle system in regions I, II, and III by simply combining
the metric of the left and the right particles, because there is
no interaction before the collision. Figure~2 shows the schematic
spacetime structure adding one dimension to Fig.~1. Our goal is to
construct an AH on the new slice, i.e., on the union of the two
null surfaces $u=0,~v>0$ and $u>0,~v=0$. By the left-right
symmetry (we work in the center-of-mass frame), it is sufficient
to consider the $u>0,~v=0$ surface. We introduce a coordinate
$\phi$ such that the metric in region II is given by
\begin{equation}
\label{metr1}
 ds^2=-dudv
+\left[1+(D-3)\frac{u}{r^{D-2}}\right]^2dr^2 +r^2\left[1-
\frac{u}{r^{D-2}}\right]^2\left(d\phi^2+\sin^2\phi
d\Omega^2_{D-4}\right).
\end{equation}
The radial coordinate $r$ in region II is adapted to the left particle,
which is thus located at $r=0$. In these coordinates, the right particle
will cross the transverse collision plane $u=v=0$ at a point distance
$b$ from the origin, where $b$ is the impact parameter. We will choose
coordinate $\phi$ so that this point is $r=b$, $\phi=0$. This setup is
identical to the one used in \cite{EG02} and \cite{YN03}.

\subsection{AH equation and boundary conditions}

\begin{figure}[tb]
\centering
{
\includegraphics[width=0.3\textwidth]{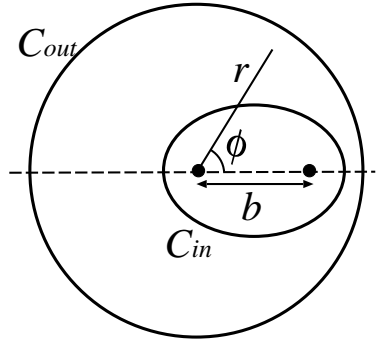}
} \caption{Schematic picture of the boundaries $C_{in}$ and
$C_{out}$. We should solve for $h(r,\phi)$ in the region
surrounded by the two boundaries.  }
\end{figure}

The schematic shape of the AH on the new slice is also shown in
Fig.~2. Because $u=r^{D-2}$ is a coordinate singularity, we have
two boundaries in this analysis: $C_{in}$ at $u=v=0$ and $C_{out}$
at $u=r^{D-2}, v=0$. We show the schematic shapes of $C_{in}$ and
$C_{out}$ in Fig. 3. Between these boundaries, the AH shape is
specified by an unknown function $u=h(r,\phi)$. The tangent vector
$k^{\mu}$ of the null geodesic congruence of the AH surface can be
found in terms of this function using metric (\ref{metr1}) and is
given by
\begin{align}
k^{u}&=\frac12\left\{
\left[1+(D-3)\frac{h}{r^{D-2}}\right]^{-2}h_{,r}^2
+r^{-2}\left(1-\frac{h}{r^{D-2}}\right)^{-2}h_{,\phi}^2
\right\},\\
k^{v}&=2,\\
k^{r}&=\left[1+(D-3)\frac{h}{r^{D-2}}\right]^{-2}h_{,r},\\
k^{\phi}&=r^{-2}\left(1-\frac{h}{r^{D-2}}\right)^{-2}h_{,\phi}.
\end{align}
Imposing that this congruence has zero expansion, we get the AH
equation:
\begin{multline}
\left(r^{D-2}-h\right)^2
\left\{
h_{,rr}+(D-3)\frac{h_{,r}}{r}
\left[
1+\frac{(D-2)h-(3/2)rh_{,r}}{r^{D-2}+(D-3)h}+\frac{(D-2)h
-(1/2)rh_{,r}}{r^{D-2}-h}
\right]
\right\}+
\\
r^{-2}\left[r^{D-2}+(D-3)h\right]^2 \left\{
h_{,\phi\phi}+(D-4)\cot\phi h_{,\phi}+ \frac{h_{,\phi}^2}{2}
\left[ \frac{D-3}{r^{D-2}+(D-3)h}-\frac{D-7}{r^{D-2}-h} \right]
\right\}
\\
=0.
\label{AHeq}
\end{multline}

Now we consider the boundary conditions.
At $C_{in}$, the continuity of the AH requires
\begin{equation}
h(r,\phi)=0.
\label{Cin_BC_1}
\end{equation}
Another condition comes from the continuity of the null tangent vector
(up to a factor). This condition is equivalent to
$k^u(x)k^v(x^\star)=k^v(x)k^u(x^\star)$ or
\begin{equation}
\left(h_{,r}^2+r^{-2}h_{,\phi}^2\right)\big|_{x}
\left(h_{,r}^2+r^{-2}h_{,\phi}^2\right)\big|_{x^\star}=16,
\label{Cin_BC_2}
\end{equation}
where $x^\star$ denotes the point symmetric to $x$ with respect to
the center of $C_{in}$ (i.e. the point $r=b/2,\phi=0$).

Now we turn to the boundary conditions at $C_{out}$.
Because $u=r^{D-2}$ is a coordinate singularity, $C_{out}$ has to
be located at some fixed unknown radius $r=r_{max}$
so that the AH is continuous. Hence we have
\begin{equation}
h=r_{max}^{D-2}, \label{Cout_BC_1}
\end{equation}
on $C_{out}$.

The last boundary condition follows by imposing the continuity of $k^{\mu}$ on $C_{out}$.
For this purpose, we should translate $k^{\mu}$ into the
$(\bar{u}, \bar{v}, \bar{r}, \bar{\phi})$ coordinates. Using the
fact that $h$ behaves like
\begin{equation}
h=r_{max}^{D-2}+h_{,r}(\phi)(r-r_{max})+\frac12
h_{,rr}(\phi)(r-r_{max})^2+\cdots \label{series}
\end{equation}
in the neighborhood of $C_{out}$, we obtain
\begin{align}
k^{\bar{u}}&=1,\\
k^{\bar{v}}&=r_{max}^{6-2D}
-\frac{4}{(D-2)r_{max}^{D-3}}\left[h_{,r}-(D-2)r_{max}^{D-3}\right]/F,\\
k^{\bar{r}}&=-r_{max}^{3-D}+2(D-2)^{-1}h_{,r}/F,
\label{kbarr}\\
k^{\bar{\phi}}&=2r_{max}^{2D-6}\left[h_{,r}-(D-2)r_{max}^{D-3}\right]^{-2}
(r-r_{max})^{-1}h_{,r\phi}/F.
\label{kbarphi}
\end{align}
where
\begin{equation}
F=(D-2)^{-2}h_{,r}^2+r_{max}^{2D-6}\left[h_{,r}-(D-2)r_{max}^{D-3}\right]^{-2}h_{,r\phi}^2.
\end{equation}
In order that $k^{\mu}$ be continuous, $k^{\bar{v}}$ should be
constant for all $\phi$:
\begin{equation}
\left[{h_{,r}-(D-2)r_{max}^{D-3}}\right]/F
=\bar{B}.
\label{B1_condition}
\end{equation}
There are also two other conditions given by
\begin{align}
&k^{\bar{r}}=Br_{max}^{3-D}\cos\phi,
\label{condition_barr}\\
&k^{\bar{\phi}}=-\frac{Br_{max}^{3-D}}{\bar{r}}\sin\phi,
\label{condition_barphi}
\end{align}
where we have used the symmetry of this system. (These conditions
are analogous to the conditions for smoothness of a
non-axisymmetric surface written in cylindrical coordinates.)
Here, $B$ and $\bar{B}$ are related as
\begin{equation}
\bar{B}=\frac{(D-2)r_{max}^{3-D}}{4}(1-B^2)
\label{BB1_relation}
\end{equation}
by the null condition $k_{\mu}k^{\mu}=0$.  Using Eqs.
\eqref{kbarr}, \eqref{B1_condition}, \eqref{condition_barr} and
\eqref{BB1_relation}, we derive
\begin{equation}
h_{,r}=(D-2)r_{max}^{D-3}\left(1+\frac{1-B^2}{1+B^2+2B\cos\phi}\right).
\label{Cout_BC_2}
\end{equation}
This is the second boundary condition at $C_{out}$. Although we
have not used Eqs.~\eqref{kbarphi} and \eqref{condition_barphi},
we can easily check the consistency. The boundary condition
\eqref{Cout_BC_2} is also consistent with the AH
equation~\eqref{AHeq}. Substituting the series \eqref{series} into
Eq.~\eqref{AHeq}, the leading-order term is
\begin{multline}
h_{,r\phi\phi}+(D-4)\cot\phi h_{,r\phi}
+\frac{(1/2)(D-7)h_{,r\phi}^2}{h_{,r}-(D-2)r_{max}^{D-3}}
\\
+\frac{(D-3)}{2(D-2)^2}r_{max}^{6-2D}h_{,r}
\left[h_{,r}-(D-2)r_{max}^{D-3}\right]\left[h_{,r}-2(D-2)r_{max}^{D-3}\right]=0.
\label{AHeq_Cout}
\end{multline}
By substituting Eq.~\eqref{Cout_BC_2} into the left-hand side, we
can confirm that this equation is actually satisfied.

In summary, there are two boundary conditions for each boundary:
Eqs.~\eqref{Cin_BC_1} and \eqref{Cin_BC_2} for $C_{in}$ and
Eqs.~\eqref{Cout_BC_1} and \eqref{Cout_BC_2} for $C_{out}$. We
should determine the shape of the boundary $C_{in}$ and the values
of $r_{max}$ and $B$, as well as the function $h(r,\phi)$, so as
to be consistent with these four boundary conditions.

The AH equation \eqref{AHeq} is highly nonlinear and finding
analytic solutions for $b\neq 0$ seems almost impossible even for
$D=4$. This is in contrast with the old-slice case~\cite{EG02},
where the AH equation was given by the Laplace equation. But as a
numerical problem, the new case is quite similar to the old case
except that one boundary is added (the $C_{in}$ boundary
conditions are the same as in the old case). We can solve this
problem by extending the numerical techniques developed in
\cite{YN03} as explained later.

\subsection{Head-on collision case}

In the head-on collision case, we can solve the AH equation
analytically. In this case, the function $h$ depends only on $r$
and the boundaries $C_{in}$ and $C_{out}$ are given by $r=r_{min}$
and $r=r_{max}$, respectively. The equation and the boundary
conditions become
\begin{equation}
h_{,rr}+\frac{(D-3)}{r}h_{,r}
\left[
1
+\frac{(D-2)h-(3/2)rh_{,r}}{r^{D-2}+(D-3)h}
+\frac{(D-2)h-(1/2)rh_{,r}}{r^{D-2}-h}
\right]=0,
\end{equation}
\begin{align}
&h(r_{min})=0,\\
&h_{,r}(r_{min})=2,\\
&h(r_{max})=r_{max}^{D-2},\\
&h_{,r}(r_{max})=2(D-2)r_{max}^{D-3}.
\end{align}
In $D=4$ case, the solution is given by
\begin{align}
& h=2r^2\log r,\\
&r_{min}=1,\\
&r_{max}=\sqrt{e}.
\end{align}
In $D\ge 5$ cases, the solutions are as follows:
\begin{align}
&h=\frac{2}{(D-4)}r^{D-2}\left(r^{D-4}-1\right),\\
&r_{min}=1,\\
&r_{max}=\left(\frac{D-2}{2}\right)^{1/(D-4)}.
\end{align}
The total AH area is easily calculated. Restoring the length
units, we have
\begin{equation}
A_{D-2}=\frac{2\Omega_{D-3}}{(D-2)}
\left(\frac{8\pi G_D \mu}{\Omega_{D-3}}\right)^{(D-2)/(D-3)}.
\end{equation}
This is exactly the same value as the area of two Penrose flat
disks (i.e., the AH on the old slice) given in \cite{EG02}. This
coincidence can be interpreted as follows. Because the null
geodesic congruence of the Penrose flat disk does not have shear,
the expansion rate does not change (i.e.~stays equals to zero)
while propagating into region II according to Raychaudhuri's
equation. Hence, the null geodesic congruence of the AH on the new
slice $v=0, u>0$ is the same as that of the Penrose flat disk, and
the areas of the two AHs coincide.

However, for positive impact parameters (grazing collisions), the
shear of the null geodesic congruence of the AH on the old slice
will be non-zero. While propagating, the expansion rate becomes
negative according to Raychaudhuri's equation, and the null
geodesic congruence of the AH on the new slice will not be the
same as on the old slice. This suggests that, using the new slice,
we should find larger AH areas and larger maximal impact
parameters compared to those
in the previous results of \cite{EG02, YN03}.

\subsection{Numerical method for grazing collision}

In the grazing collision case, a numerical calculation is required
to determine the AH. To solve the AH equation, we introduce
coordinates $(\tilde{r}, \tilde{\phi})$ by
\begin{align}
r\cos\phi&=\tilde{r}\cos\tilde{\phi}+b/2,\\
r\sin\phi&=\tilde{r}\sin\tilde{\phi}.
\end{align}
In these coordinates, the central point of $C_{in}$ is given by
$\tilde{r}=0$. (Because of the left-right symmetry, $C_{in}$ will
be symmetric with respect to the lines $\tilde{\phi}=0$ and
$\tilde{\phi}=\pi/2$.) We specify the boundaries $C_{in}$ and
$C_{out}$ by
\begin{align}
\tilde{r}&=g_{in}(\tilde{\phi}),\\
\tilde{r}&=g_{out}(\tilde{\phi})=-(1/2)b\cos\phi+
\left[r_{max}^2-(1/4)b^2\sin^2\tilde{\phi}\right]^{1/2},
\end{align}
respectively. We make a further transformation
\begin{equation}
R=\frac{\tilde{r}-g_{in}(\tilde{\phi})}{g_{out}(\tilde{\phi})-g_{in}(\tilde{\phi})},
\end{equation}
and solve the AH equation using the $(R,\tilde\phi)$ coordinates.
The advantage of these coordinates is that $C_{in}$ and $C_{out}$
are specified by $R=0$ and $R=1$, respectively, and the boundary
conditions are easily imposed.

The following algorithm was used in the numerical solution of our
boundary value problem. First, some test boundaries $C_{in}$ and
$C_{out}$ are taken, and the solution of the AH equation
satisfying only the two boundary conditions \eqref{Cin_BC_1} and
\eqref{Cout_BC_1} is constructed (using a finite-difference
discretization of the partial differential equation, and a
convergent iterative procedure). Next, the difference from the
boundary condition \eqref{Cin_BC_2} is calculated:
\begin{equation}
\Delta_{in}(\tilde{\phi})=
\left(h_{,r}^2+r^{-2}h_{,\phi}^2\right)\big|_{x}
\left(h_{,r}^2+r^{-2}h_{,\phi}^2\right)\big|_{x^\star}-16,
\end{equation}
and $C_{in}$ is modified as follows:
\begin{equation}
g_{in}^{\rm next}(\tilde{\phi})=g_{in}(\tilde{\phi})+\epsilon_{in}\Delta_{in}(\tilde{\phi}).
\end{equation}
The $C_{out}$ is also modified at this step, as follows. Recall
that $C_{out}$ is characterized by $r_{max}$ and $B$. We determine
$B$ using Eq.~\eqref{Cout_BC_2} at $\tilde{\phi}=0$ and calculate
the difference from the boundary condition \eqref{Cout_BC_2} at
$\tilde{\phi}=\pi$:
\begin{equation}
\Delta_{out}=
h_{,r}-2(D-2)r_{max}^{D-3}/{(1-B)}.
\end{equation}
Using this value, we modify $r_{max}$ as follows:
\begin{equation}
r_{max}^{\rm next}=r_{max}+\epsilon_{out}\Delta_{out}.
\end{equation}
If we choose $\epsilon_{in}$ and $\epsilon_{out}$ appropriately,
we can make the boundaries converge by iterating these steps.
We truncated the iteration when the absolute values of $\Delta_{in}(\tilde{\phi})$
and $\Delta_{out}$ become less than $10^{-5}$.

\begin{table}[tb]
\caption{The error estimated by the difference from the $C_{out}$
boundary condition, i.e., Eq.~\eqref{averaged_error} evaluated at
$b=b_{max}$. The error decreases by a factor of about $4$ when
doubling the resolution. This indicates the correctness of our
numerical calculations. }
\begin{ruledtabular}
\begin{tabular}{c|cccccccc}
 $D$ & $4$ & $5$ &$6$ & $7$ & $8$ & $9$ & $10$ & $11$ \\
  \hline
$(25\times 50)$ grids & $0.21\%$ & $2.0\%$ & $4.2\%$ & $-$ & $-$ & $-$ & $-$ & $-$ \\
$(50\times 100)$ grids & $0.052\%$ & $0.49\%$ &$1.1\%$ & $1.9\%$ & $3.0\%$ & $4.0\%$ & $4.6\%$ & $5.1\%$ \\
$(100\times 200)$ grids & $-$ & $-$ &$0.29\%$ & $0.50\%$ & $0.75\%$ & $1.1\%$ & $1.4\%$ & $1.8\%$ \\
\end{tabular}
\end{ruledtabular}
\end{table}

To evaluate the numerical error, the following method was used. In
the numerical method explained above, we could use only two grid
points to impose the boundary condition at $C_{out}$, because the
values to be determined are only $r_{max}$ and $B$. In principle,
the boundary conditions~\eqref{Cout_BC_2} should be satisfied at
the remaining grid points because Eq.~\eqref{Cout_BC_2} is
compatible with the AH equation in the neighborhood of $C_{out}$,
as expressed by~\eqref{AHeq_Cout}. However, in practice, because
of the finiteness of the number of grid points, small differences
from the boundary condition~\eqref{Cout_BC_2} will be present in
the intermediate grid points ($0<\phi<\pi$). This suggests to
estimate the characteristic error as follows:
\begin{equation}
\delta=\frac{1}{N}\sum \left|1-h_{,r}^{-1}(D-2)r_{max}^{D-3}\left(
1+\frac{1-B^2}{1+B^2+2B\cos\phi}\right)\right|,
\label{averaged_error}
\end{equation}
where the sum is taken over all grid points on $C_{out}$, and $N$
is the total number of these points.

Table I summarizes the resolution of the grids used to discretize
the $(R,\tilde{\phi})$ coordinates in our computations, as well as
the error $\delta$ at $b=b_{max}$. We observed that the error
estimated by $\delta$ becomes larger as $b$ increases, and takes
the largest value at $b=b_{max}$; it also becomes larger for
larger $D$. For fixed $b$ and $D$, the $\delta$ typically
decreases by a factor of about $4$ if a grid with double
resolution (i.e. with 4 times as many points) is used. Such
behavior of the error strongly indicates the correctness of our
numerical program, and the convergence to the continuum limit.
Although $\delta$ seems somewhat large for large $D$, it turned
out that the error in $b_{max}$ is much smaller, as follows by
comparing the values of $b_{max}$ obtained for different grid
resolutions. We estimate the error in $b_{max}$ at the level of
about $0.2\%$ for all $D$. Instead, the $\delta$ reflects the
error in the AH shape. To summarize, the error in
Figs.~\ref{shape_D4}-\ref{r_min},
\ref{Mlb_MAH}-\ref{MJ_allowed_D9}
 is roughly of the magnitude given in Table~I,
while the error in Table~II is about 0.2\%.

\section{Numerical results}

In this section, we present the numerical results for the AHs in
the grazing collision. The section is divided into two parts. We
first provide the results for the AH shape and the maximal impact
parameter of the AH formation. Next, we introduce two quantities
that indicate the amount of energy trapped by the AH and discuss
the final state of the produced black hole, as allowed or
prohibited by the area theorem.

\subsection{AH shape and the maximal impact parameter}

\begin{figure}[tb]
\centering
{
\includegraphics[width=0.23\textwidth]{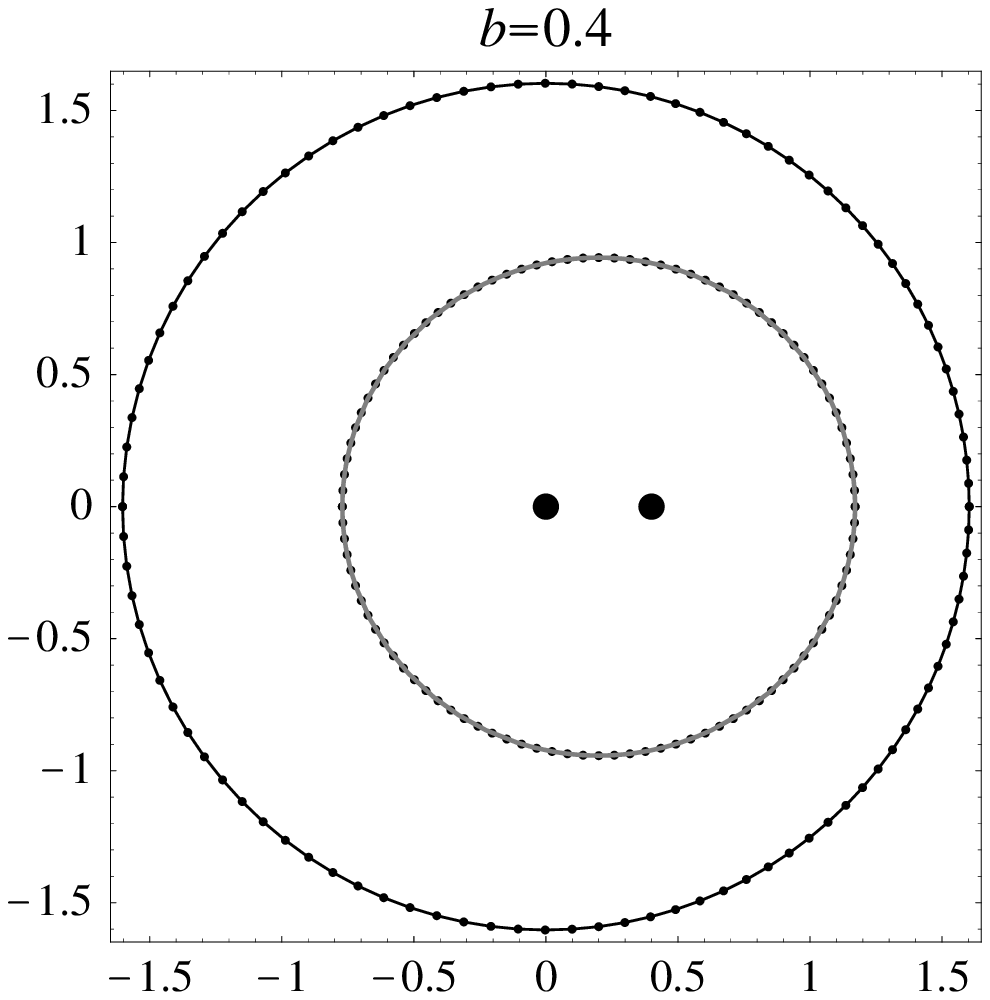}
\hspace{1mm}
\includegraphics[width=0.23\textwidth]{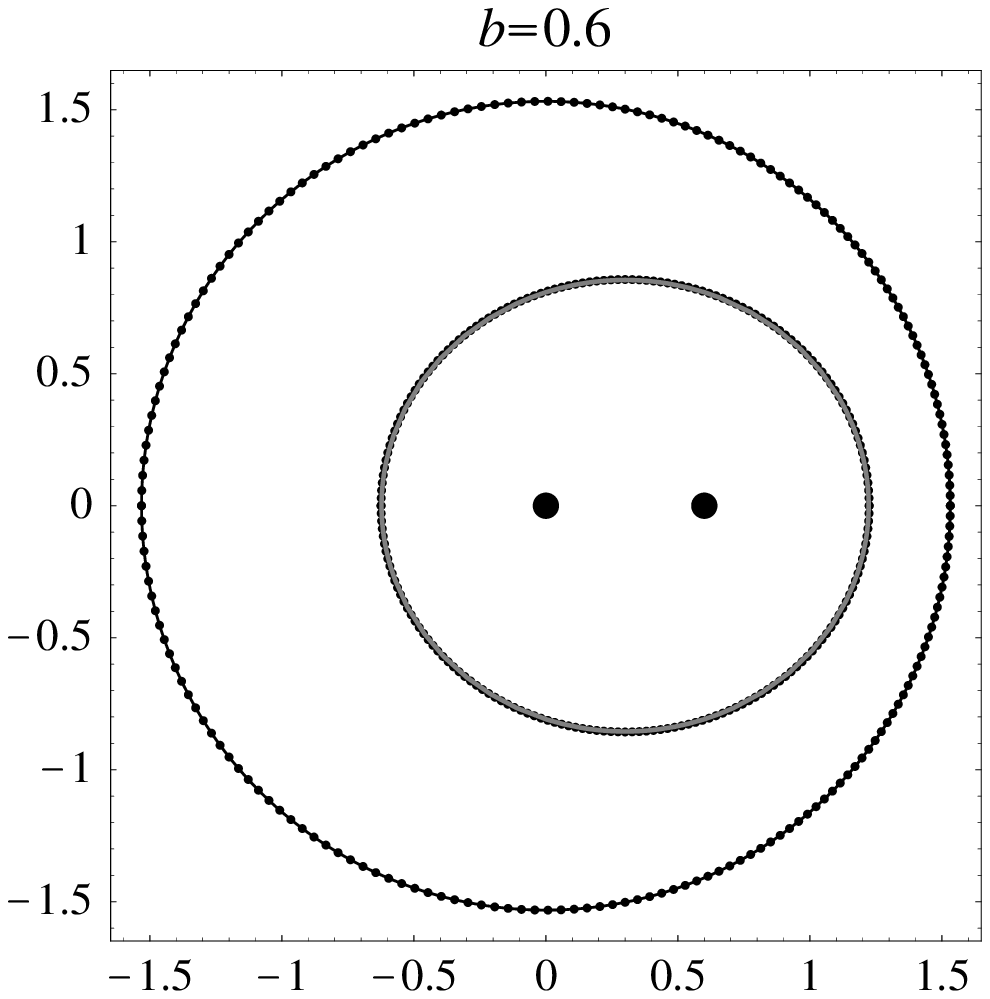}
\hspace{1mm}
\includegraphics[width=0.23\textwidth]{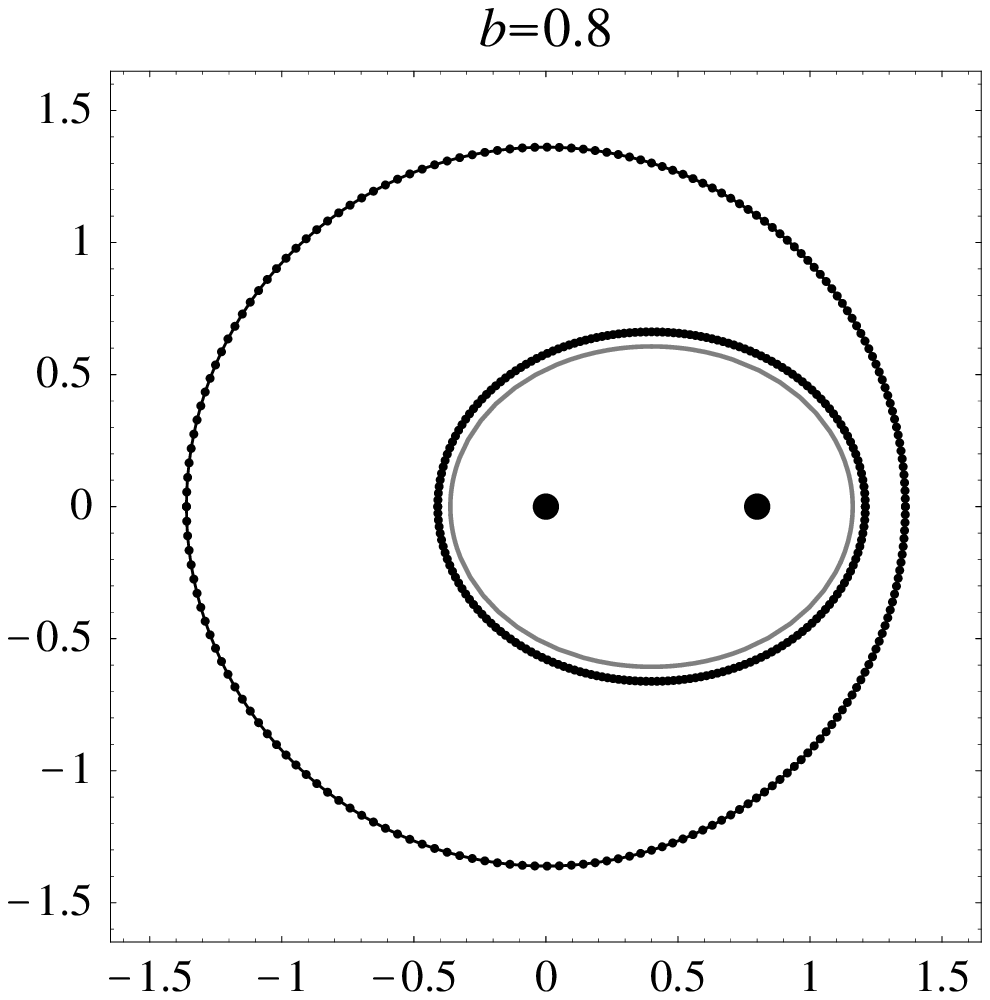}
\hspace{1mm}
\includegraphics[width=0.23\textwidth]{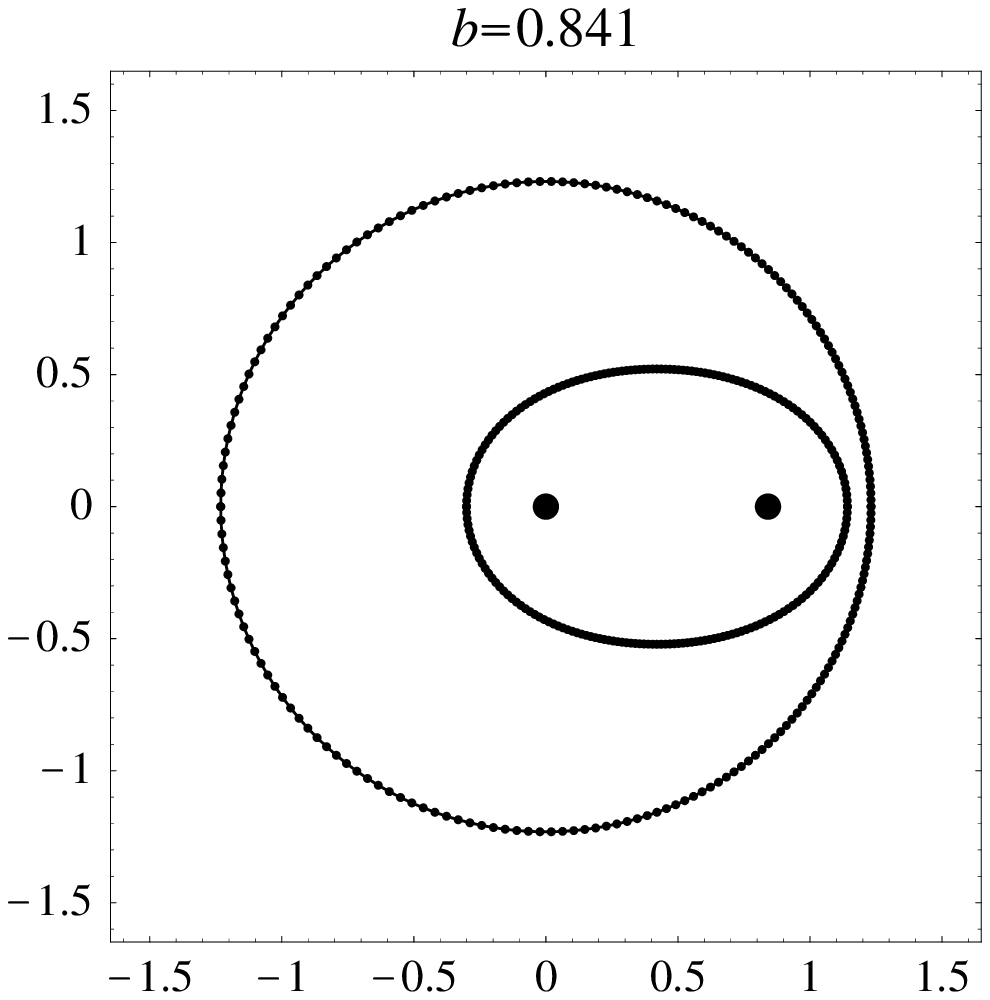}
}
\caption{The shape of $C_{in}$ and $C_{out}$ for $b/r_0=0.4, 0.6,0.8,0.841$
in the $D=4$ case.
Two dots in each figure indicate the location of incoming particles.
The old-slice AHs in~\cite{EG02} for $b/r_0=0.4,0.6,0.8$
are shown by gray lines. We could not find the AH for $b/r_0=0.842$. }
\label{shape_D4}
\end{figure}

Figure~\ref{shape_D4} shows the shapes of $C_{in}$ and $C_{out}$
for various values of $b$ in the $D=4$ case. The old-slice AHs of
Eardley and Giddings~\cite{EG02} are also shown. As $b$ increases,
$C_{in}$ becomes oblate, and $r_{max}$ becomes smaller. For small
$b$, $C_{in}$ and the corresponding boundary curve of the
old-slice AH almost coincide, which again indicates the
correctness of our numerical program. For larger $b$, $C_{in}$
lies outside the old-slice AH curve. For even larger $b$, we have
a situation when there exists an AH in the new slice, while there
is no AH in the old slice. The $b_{max}$ becomes about 5\% larger
than the previous result of~\cite{EG02}.

\begin{figure}[tb]
\centering
{
\includegraphics[width=0.23\textwidth]{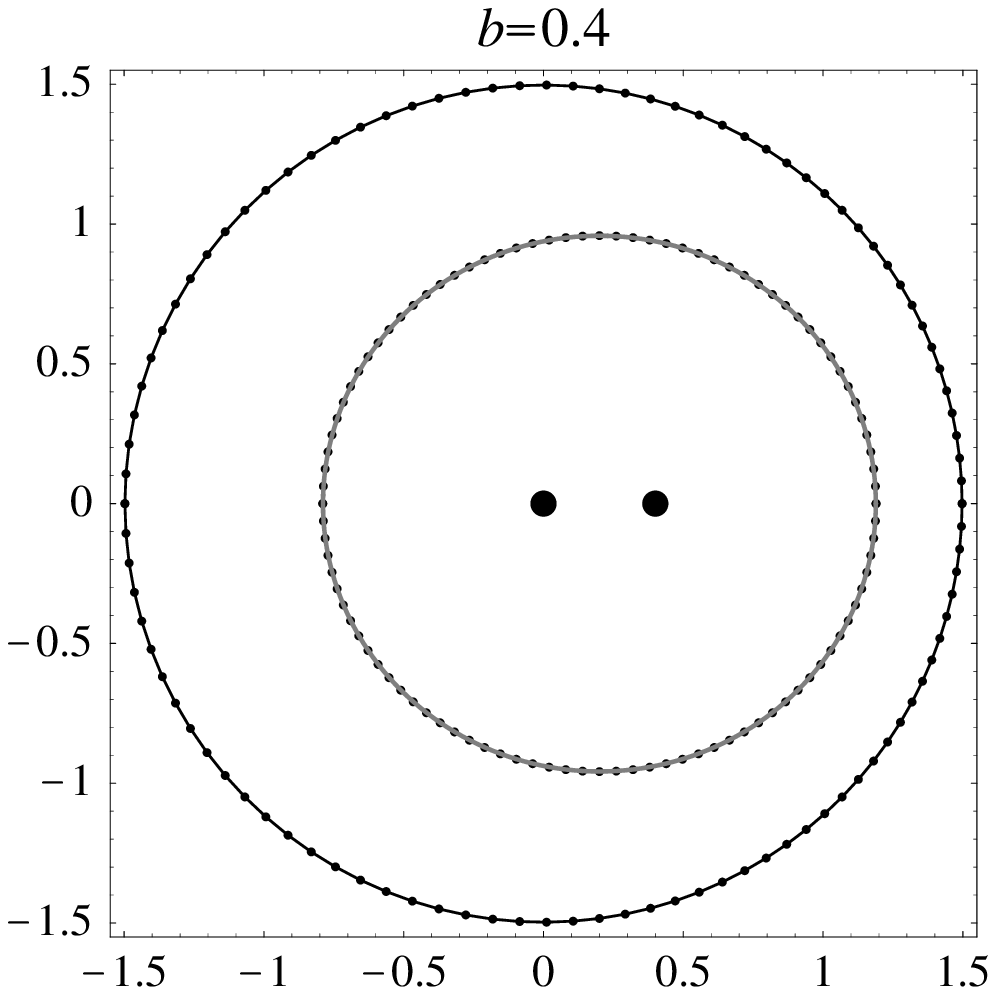}
\hspace{1mm}
\includegraphics[width=0.23\textwidth]{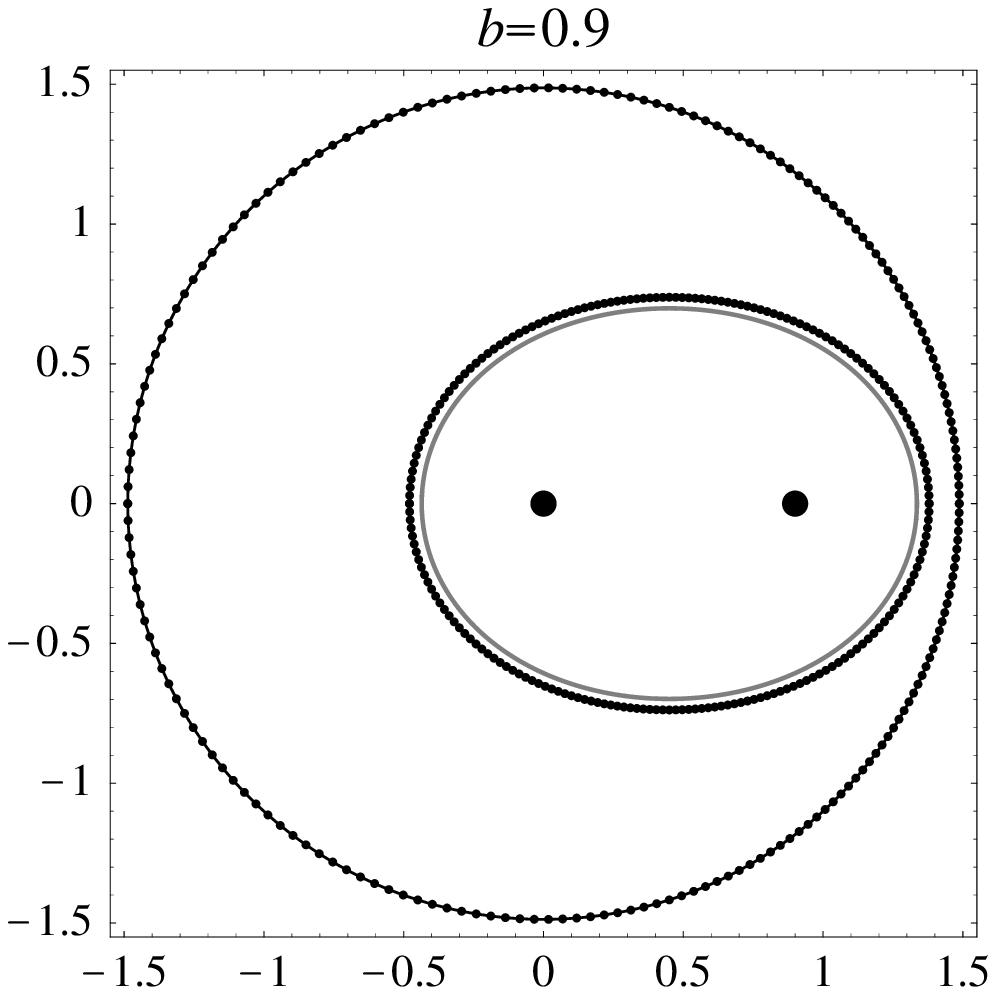}
\hspace{1mm}
\includegraphics[width=0.23\textwidth]{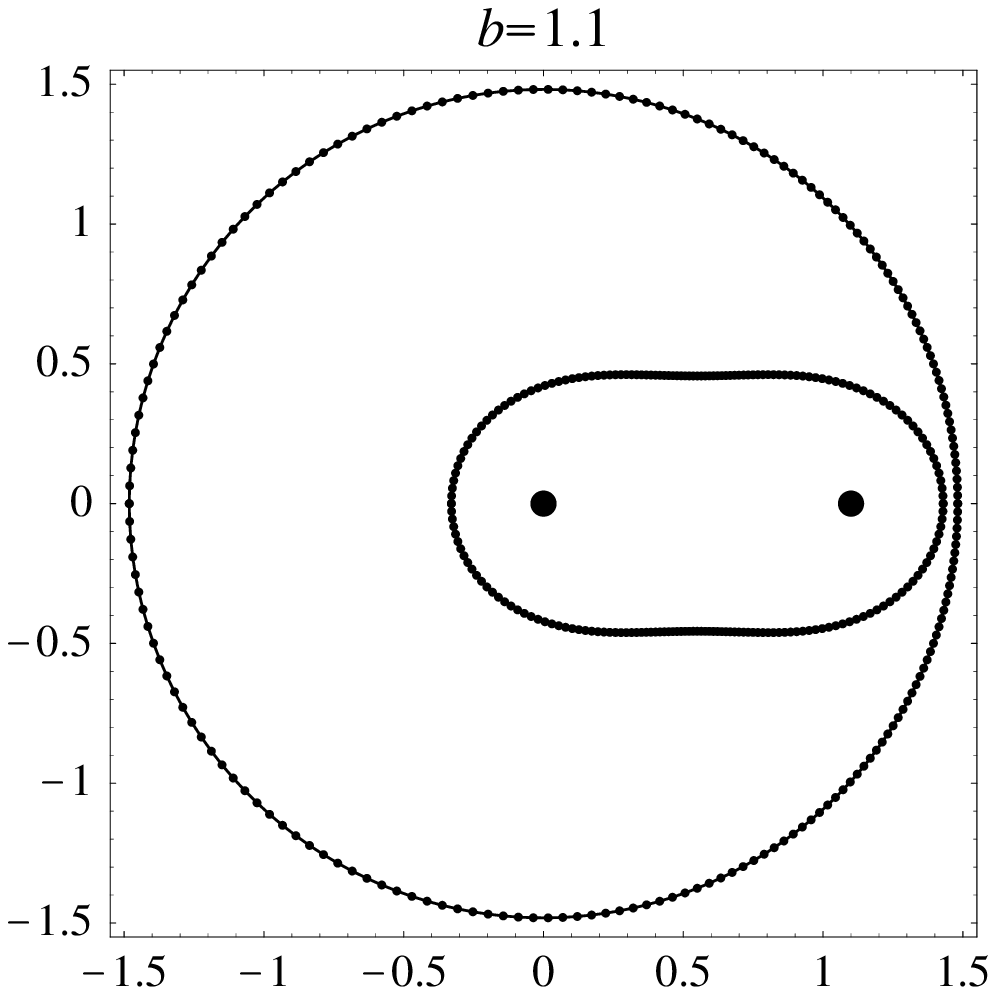}
\hspace{1mm}
\includegraphics[width=0.23\textwidth]{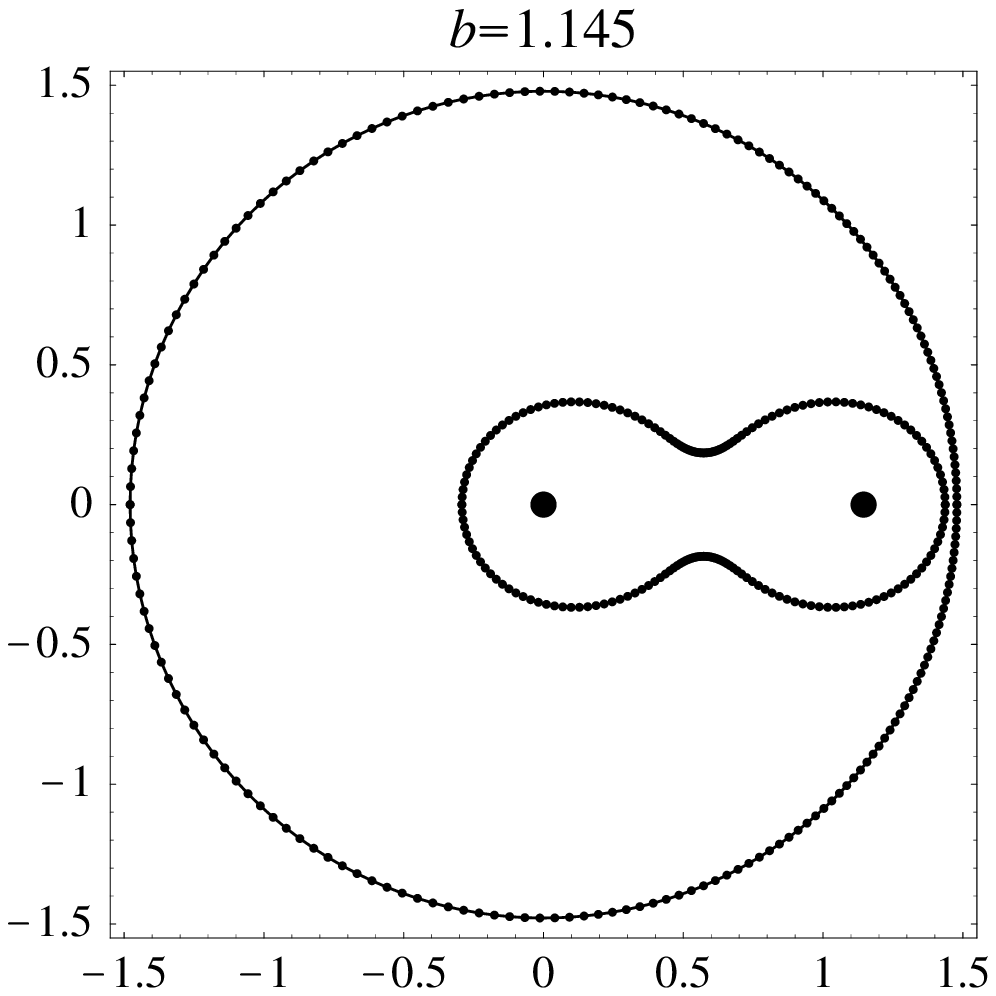}
}
\caption{The shapes of $C_{in}$ and $C_{out}$ for $b=0.4, 0.9,1.1,1.145$ in the $D=5$ case.
Two dots in each figure indicate the location of the incoming particles.  The old-slice
AHs of \cite{YN03} for $b/r_0=0.4, 0.9$ are shown by gray lines.
We could not find the AH for $b/r_0=1.146$.}
\label{shape_D5}
\end{figure}

Figure~\ref{shape_D5} shows the shapes of $C_{in}$ and $C_{out}$
in the $D=5$ case. In this case, $r_{max}$ is almost constant for
all $b$ and the shape of $C_{in}$ becomes oblate as $b$ increases.
It is quite interesting that $C_{in}$ becomes non-convex around
$b=b_{max}$. The value of $b_{max}$ is about 18\% larger than the
previous result of Yoshino and Nambu \cite{YN03}, which leads to
40\% larger cross section of the AH formation, the present value
being $\sigma_{\rm AH}\simeq 1.5\pi \left[r_h(2\mu)\right]^2$.

\begin{figure}[tb]
\centering
{
\includegraphics[width=0.23\textwidth]{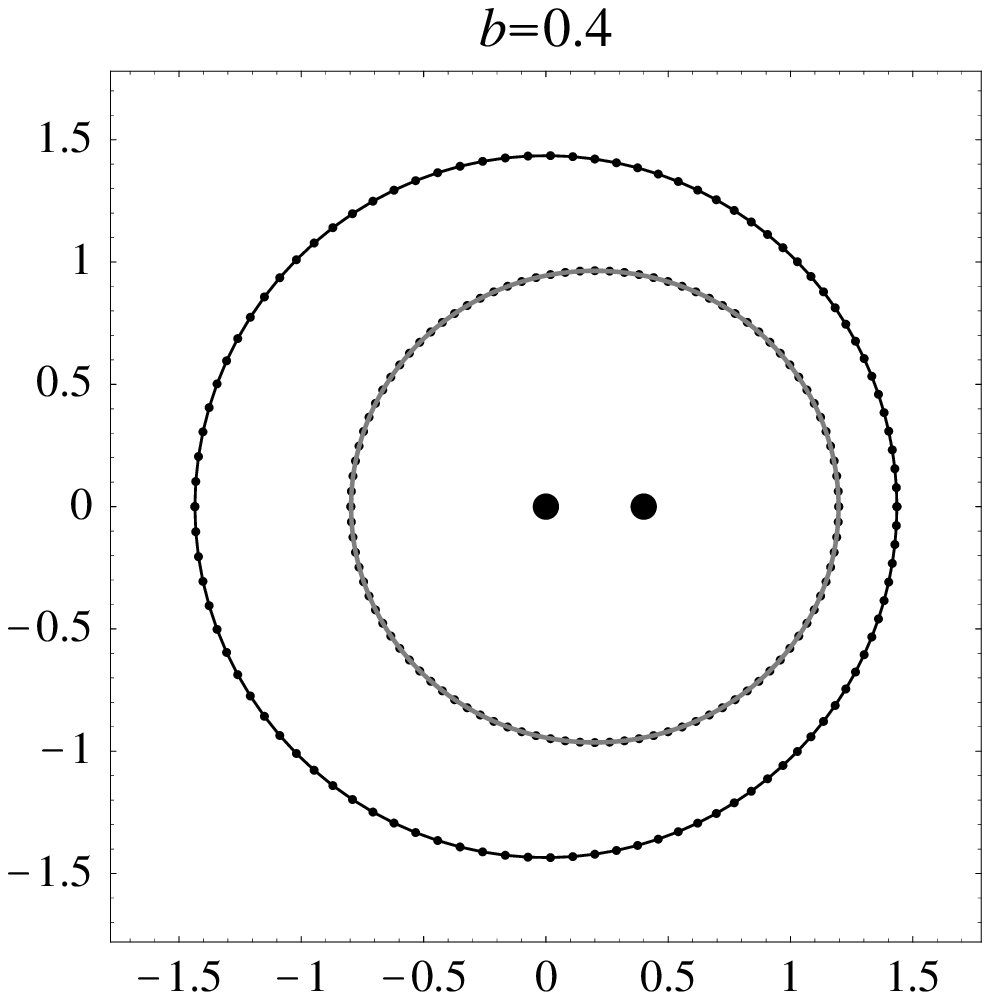}
\hspace{1mm}
\includegraphics[width=0.23\textwidth]{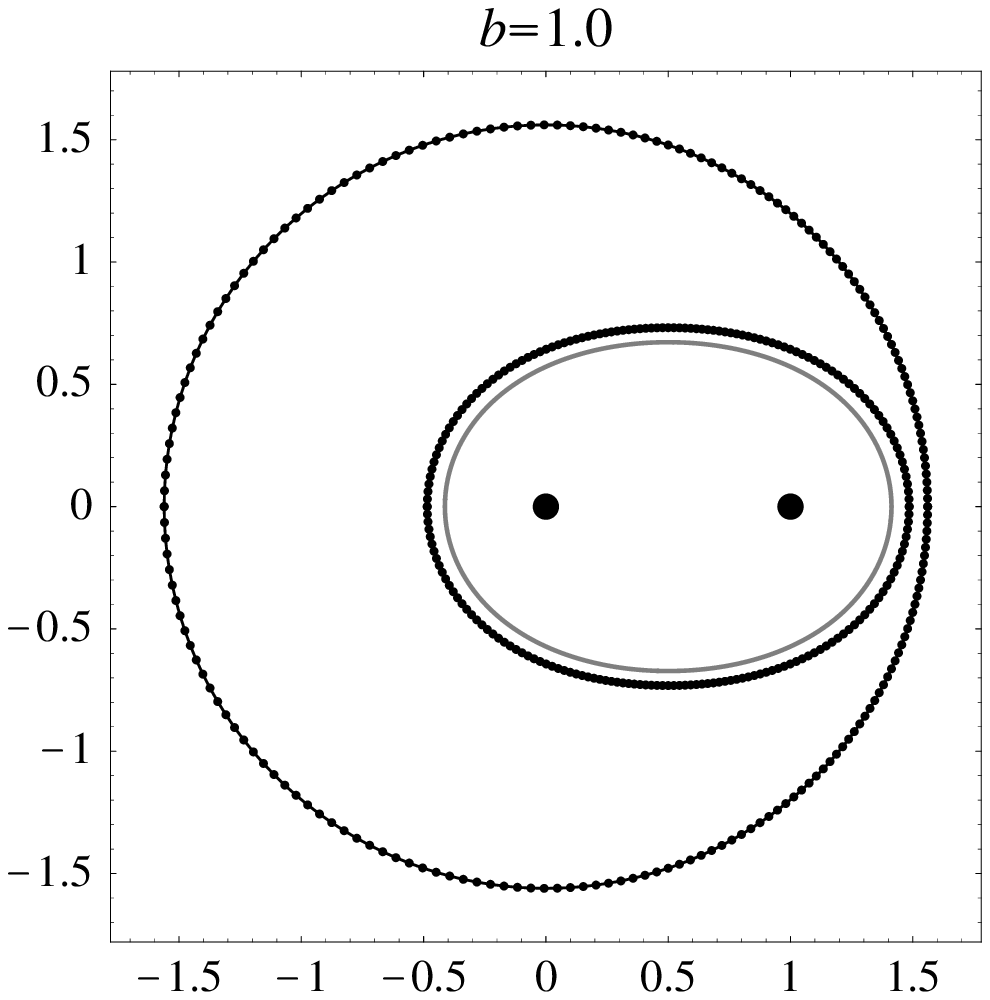}
\hspace{1mm}
\includegraphics[width=0.23\textwidth]{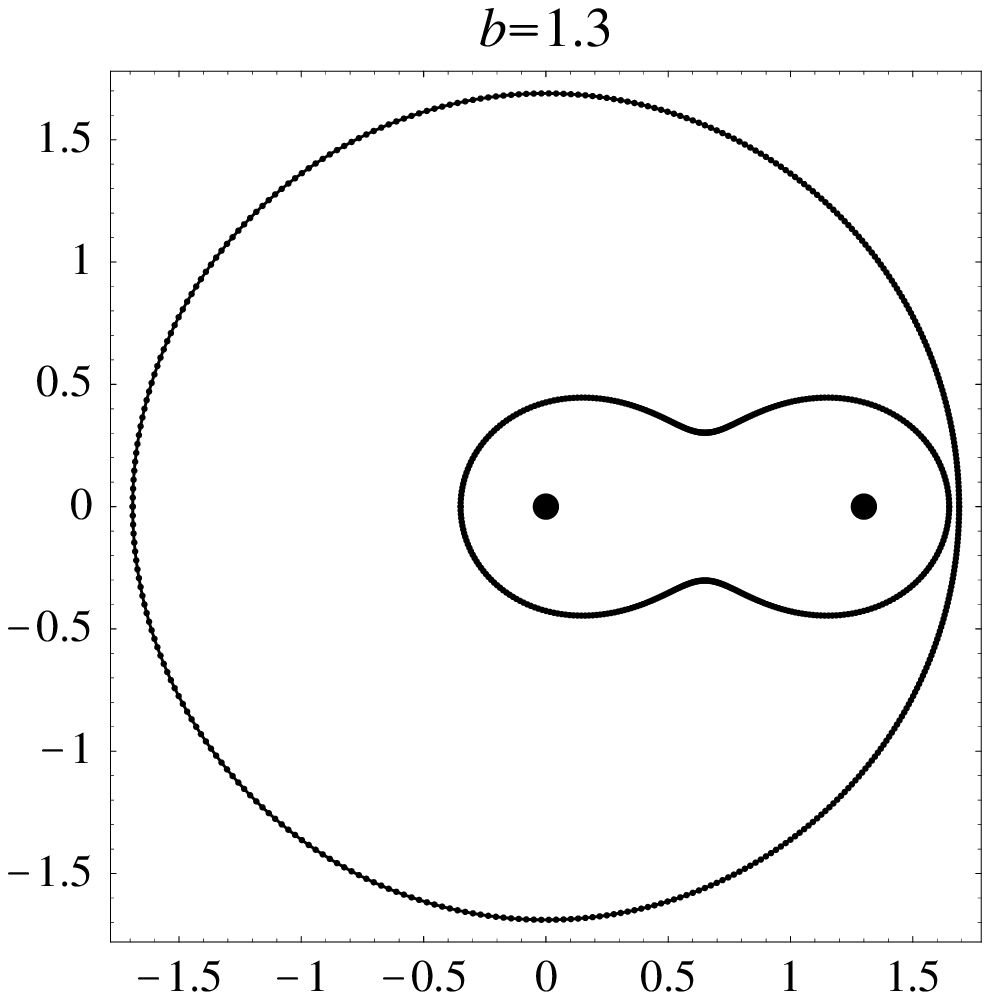}
\hspace{1mm}
\includegraphics[width=0.23\textwidth]{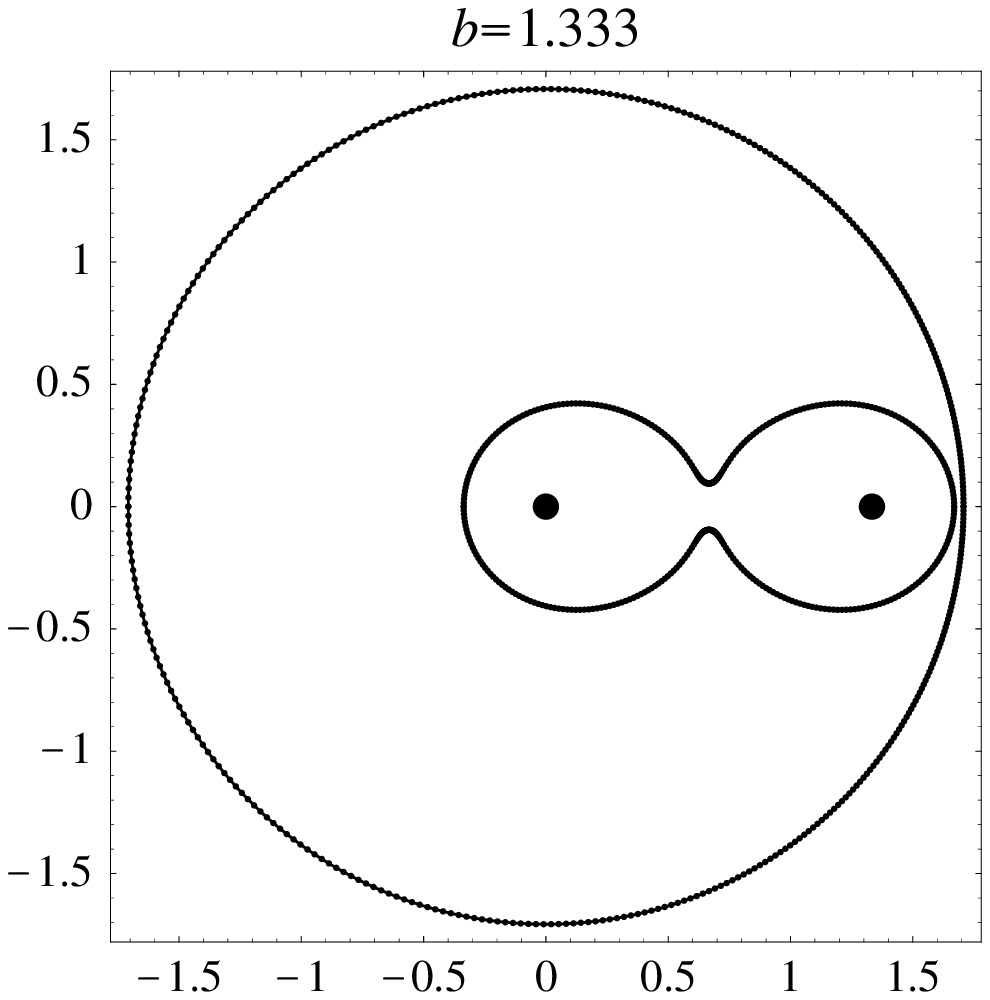}
}
\caption{The shape of $C_{in}$ and $C_{out}$ for $b/r_0=0.4, 1.0,1.3,1.333$ in the $D=6$ case.
Two dots in each figure indicate the location of the incoming particles.  The old-slice
AHs of \cite{YN03} for $b=0.4,1.0$ are shown by gray lines.
We could not find the AH for $b/r_0=1.334$.}
\label{shape_D6}
\end{figure}

Figure \ref{shape_D6} shows the shapes of $C_{in}$ and $C_{out}$
in the $D=6$ case. In this case, $r_{max}$ becomes larger as $b$
increases. The shape of $C_{in}$ at $b=b_{max}$ is even more
non-convex than that in the $D=5$ case. The value of $b_{max}$ is
about 26\% larger than the previous result of \cite{YN03}. This
leads to 59\% larger cross section of the AH formation, the
present value being $\sigma_{\rm AH}\simeq 2.1\pi
\left[r_h(2\mu)\right]^2$.

\begin{figure}[tb]
\centering
{
\includegraphics[width=0.25\textwidth]{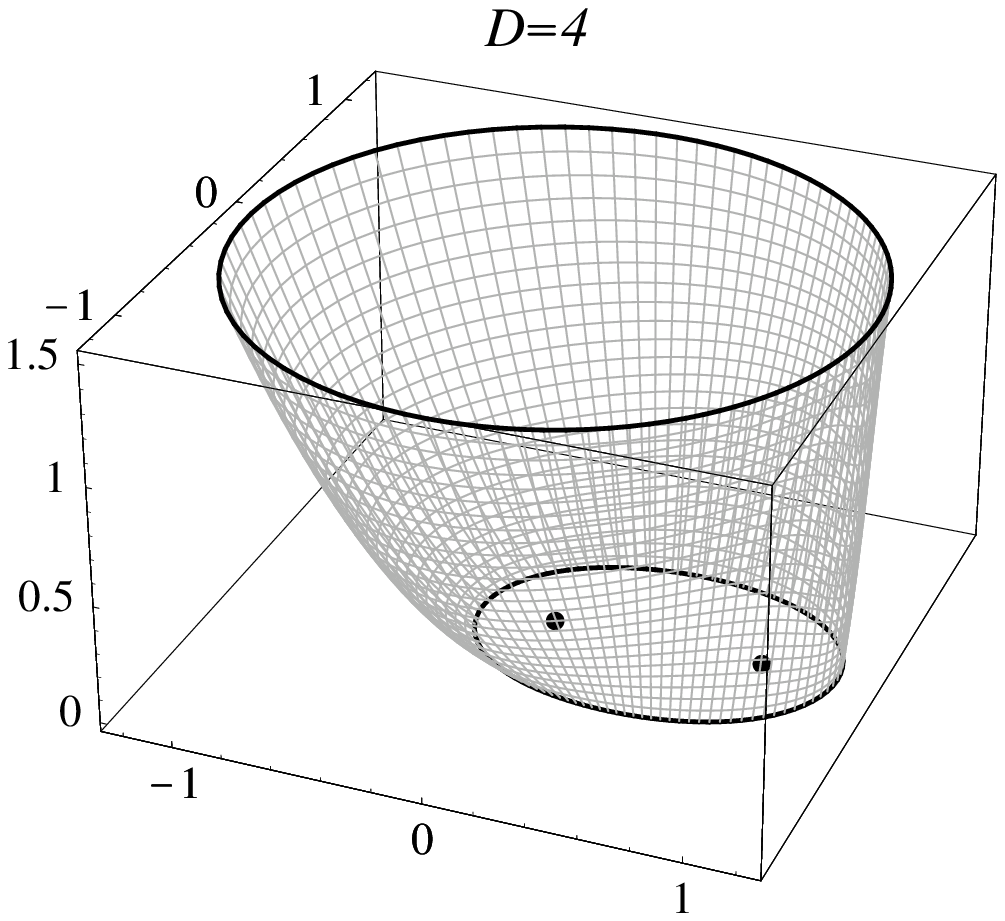}
\hspace{3mm}
\includegraphics[width=0.25\textwidth]{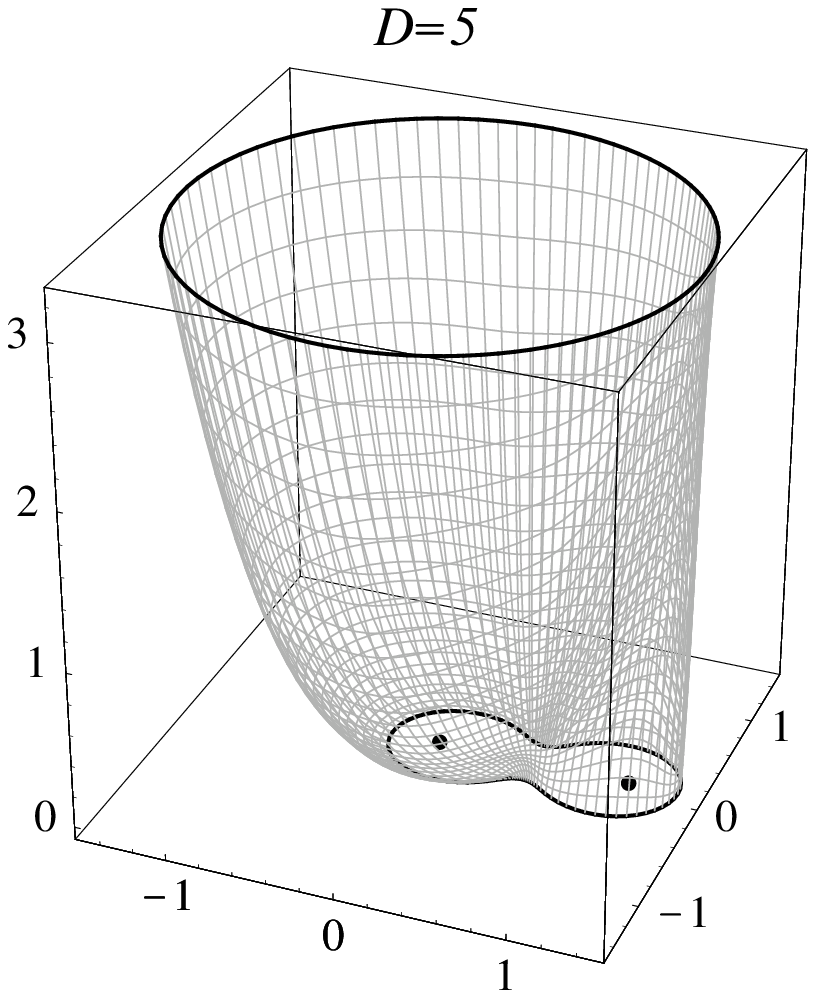}
\vspace{3mm}
\includegraphics[width=0.25\textwidth]{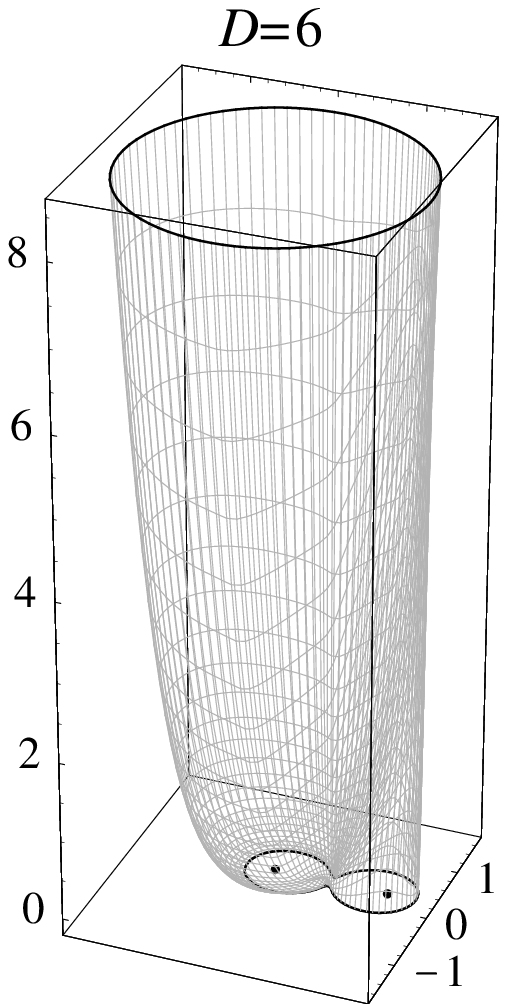}
} \caption{Three-dimensional plots of the AH shape function at
$b=b_{max}$ for $D=4,5,$ and $6$. The AH becomes taller for larger
$D$ as a consequence of the growing power exponent in the boundary
condition (\ref{Cout_BC_1}).}
\label{3D_figure}
\end{figure}

In Fig. \ref{3D_figure}, we plot the AH shape function $h(r,\phi)$
at $b=b_{max}$ for $D=4,5,$ and $6$.
The AH becomes taller for larger $D$ as a consequence of the
growing power exponent in the boundary condition
(\ref{Cout_BC_1}).

For $D\ge 7$, the shapes of $C_{in}$ and $C_{out}$ and the horizon
shape behave qualitatively similarly to the $D=6$ case, and we do
not present them here in detail.

\begin{figure}[tb]
\centering
{
\includegraphics[width=0.5\textwidth]{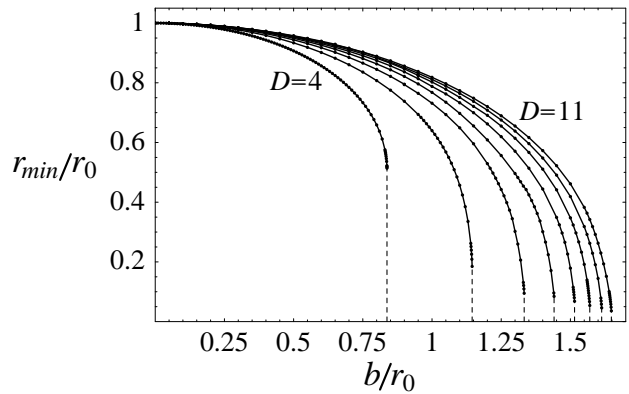}
}
\caption{The relation between $b/r_0$ and $r_{min}/r_0$, where $r_{min}=g_{in}(\pi/2)$.
The maximal impact parameter occurs at $dr_{min}/db=-\infty$.  }
\label{r_min}
\end{figure}

\begin{table}[tb]
\caption{The value of $b_{max}/r_0$, the ratio of the increase in
the maximal impact parameter, and the value of the AH formation
cross section $\sigma_{\rm AH}$ (which provides a rigorous lower
bound for $\sigma_{\rm BH}$). }
\begin{ruledtabular}
\begin{tabular}{c|cccccccc}
 $D$ & $4$ & $5$ &$6$ & $7$ & $8$ & $9$ & $10$ & $11$ \\
  \hline
 $b_{max}/r_0$& $0.841$ & $1.145$ &$1.333$ & $1.441$ & $1.515$ & $1.570$ & $1.613$ & $1.648$ \\
  $b_{max}/\hat{b}_{max}-1$ & $5\%$ & $18\%$ &$26\%$ & $29\%$ & $29\%$ & $30\%$ & $30\%$ & $30\%$ \\
 $\sigma_{\rm AH}/\pi \left[r_h(2\mu)\right]^2$ & $0.71$ & $1.54$ &$2.15$ & $2.52$ & $2.77$ & $2.95$ & $3.09$ & $3.20$ \\
  \end{tabular}
\end{ruledtabular}
\end{table}

\begin{figure}[tb]
\centering
{
\includegraphics[width=0.5\textwidth]{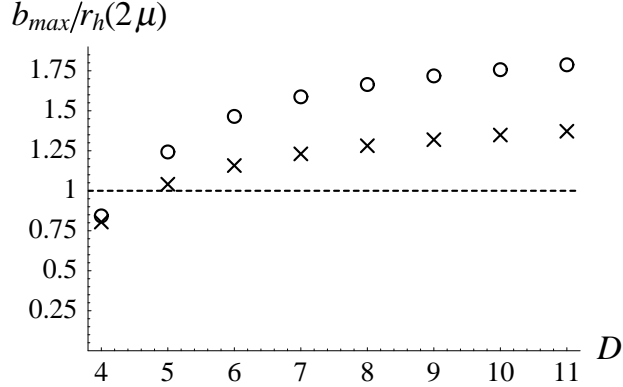}
}
\caption{Summary of the $b_{max}/r_h(2\mu)$ values ($\circ$)
and the previous $\hat{b}_{max}/r_h(2\mu)$ values ($\times$)
for $D=4,...,11$. }
\label{bmax_summary}
\end{figure}

Figure~\ref{r_min} shows the minimum radius of $C_{in}$,
$r_{min}\equiv g_{in}(\pi/2)$, as a function of $b$ for $D=4,...,11$.
The maximal impact parameter occurs at $dr_{min}/db=-\infty$. From
this figure, we can read off the values of $b_{max}$.

Table II summarizes the numerical results for $b_{max}$, which are
the most important output of our analysis. For $5\le D\le 11$, the
values of $b_{max}$ increase by $18$-$30$\% compared to the
previous values of $\hat{b}_{max}$ in \cite{YN03}.
Correspondingly, the values of the cross section of the AH
formation $\sigma_{\rm AH}/\pi \left[r_h(2\mu)\right]^2$ increase
by $40$-$70$ \%. This indicates that the black hole production
rate in accelerators can be quite a bit larger compared to the
previous estimates, this tendency being especially enhanced for
larger $D$. In the $D=4$ case, which may have only astrophysical
applications, we find only a modest 5\% improvement in $b_{max}$
compared to~\cite{EG02}. We compare the present values of
$b_{max}/r_h(2\mu)$ with the previous values of
$\hat{b}_{max}/r_h(2\mu)$ in Figure~\ref{bmax_summary}.

\subsection{Trapped energy and final state of the produced black hole}

Assuming the cosmic censorship~\cite{P69}, an event horizon (EH) must be
present outside the found AHs (see \cite{W84} for a proof).
Moreover, the area theorem~\cite{H71}
states that the EH area never decreases. Hence one naturally expects that
the AH mass defined by
\begin{equation}
M_{\rm AH}=\frac{(D-2)\Omega_{D-2}}{16\pi G_D}
\left(\frac{A_{D-2}}{\Omega_{D-2}}\right)^{(D-3)/(D-2)}
\label{AH_mass}
\end{equation}
provides the lower bound of the irreducible mass $M_{irr}$ of the
final Kerr black hole. We should mention that for this statement
to be rigorously justified, the area of an arbitrary surface
outside of the AH in a given slice should be larger than that of
the AH. In the old slice
analyzed in \cite{EG02, YN03} this property actually holds. On the
other hand, on the new slice analyzed here we can find a
counterexample. In the head-on collision case, for example, the
union of two surfaces $r=r_{c}>r_{max}, 0\le u\le r_c^{D-2}, v=0$
and $r=r_{c}, 0\le v\le r_c^{D-2}, u=0$ is closed, lies outside of
the AH, and has zero area. Thus there is no rigorous reason why in
the present analysis $M_{\rm AH}$ should provide the lower bound
on the final mass.

Nevertheless, we can still find a rigorous lower bound on
$M_{irr}$, arguing as follows. The intersection of the AH and
$u=v=0$ is given by $C_{in}$. Let us denote the intersection of
the EH and $u=v=0$ by $C_{\rm EH}$. This curve must lie outside
$C_{in}$. Further, one can show that the area $A_{\rm EH}$ of the
intersection of the EH with the old slice is equal to twice the
area of the region surrounded by $C_{\rm EH}$ calculated with the
$(D-2)$-dimensional flat metric. It follows that $A_{\rm EH}$ is
bounded below by $A_{lb}$, the latter quantity being defined as
twice the area of the region surrounded by $C_{in}$, calculated
with the same flat metric. Hence the rigorous lower bound on
$M_{irr}$ is given by
\begin{equation}
M_{lb}=\frac{(D-2)\Omega_{D-2}}{16\pi G_D}
\left(\frac{A_{lb}}{\Omega_{D-2}}\right)^{(D-3)/(D-2)}.
\label{lower_bound}
\end{equation}

\begin{figure}[tb]
\centering
{
\includegraphics[width=0.3\textwidth]{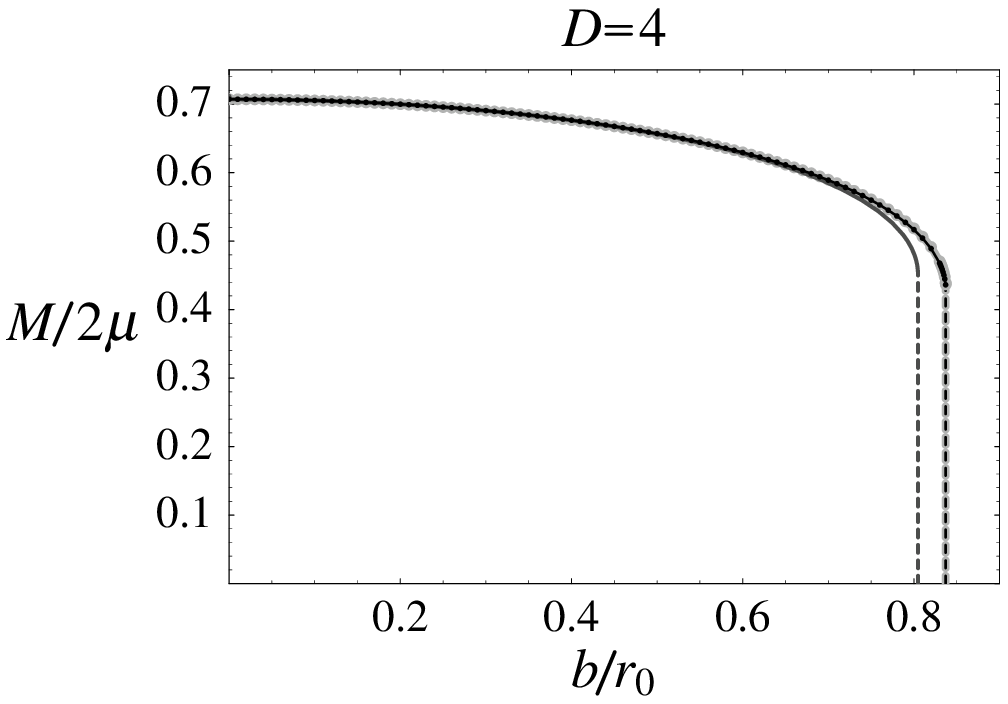}\hspace{10mm}
\includegraphics[width=0.3\textwidth]{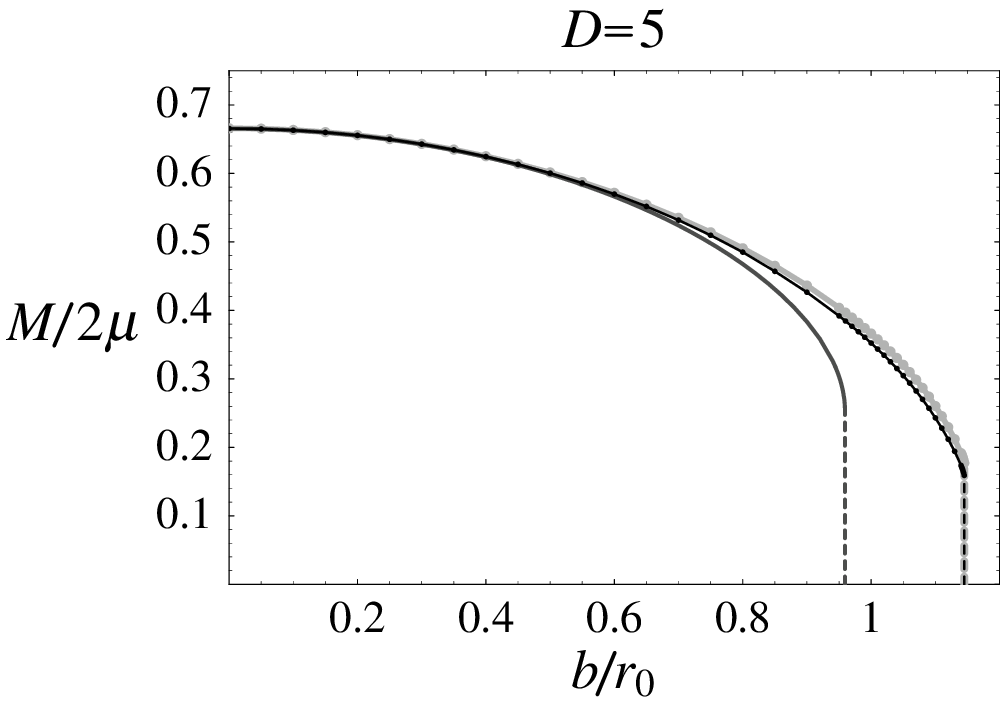}
\\
\vspace{3mm}
\includegraphics[width=0.3\textwidth]{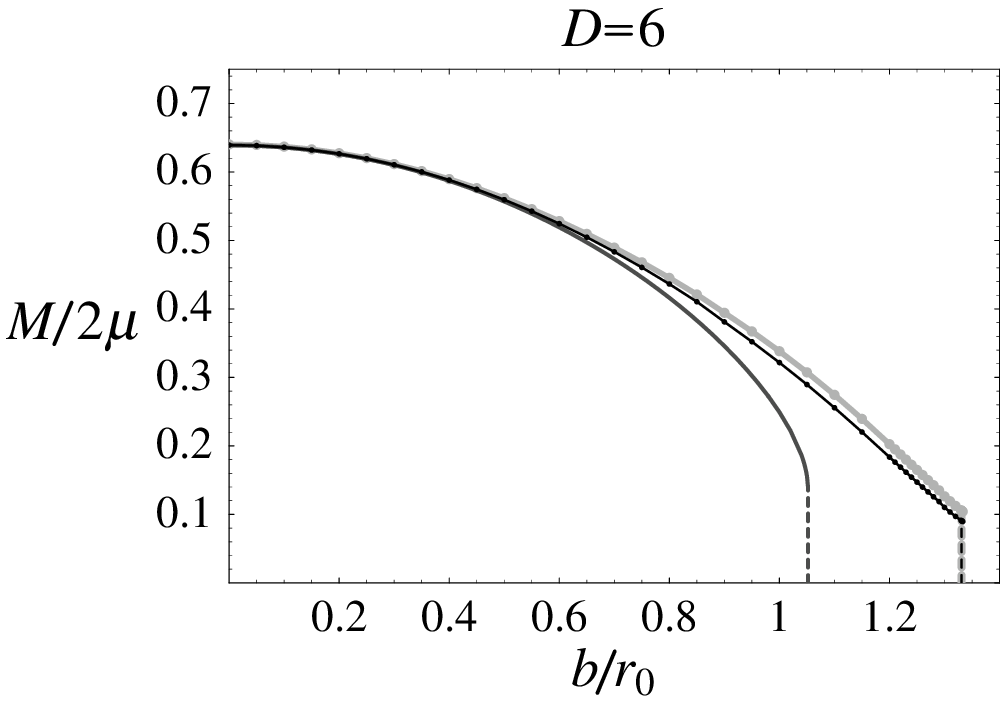}\hspace{10mm}
\includegraphics[width=0.3\textwidth]{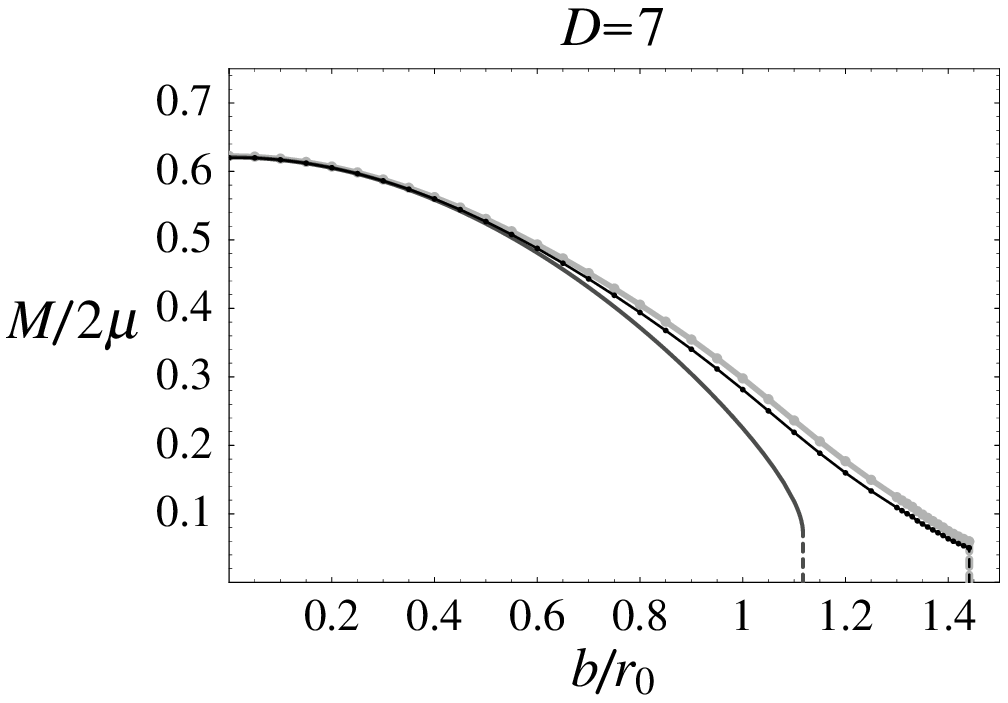}
\\
\vspace{3mm}
\includegraphics[width=0.3\textwidth]{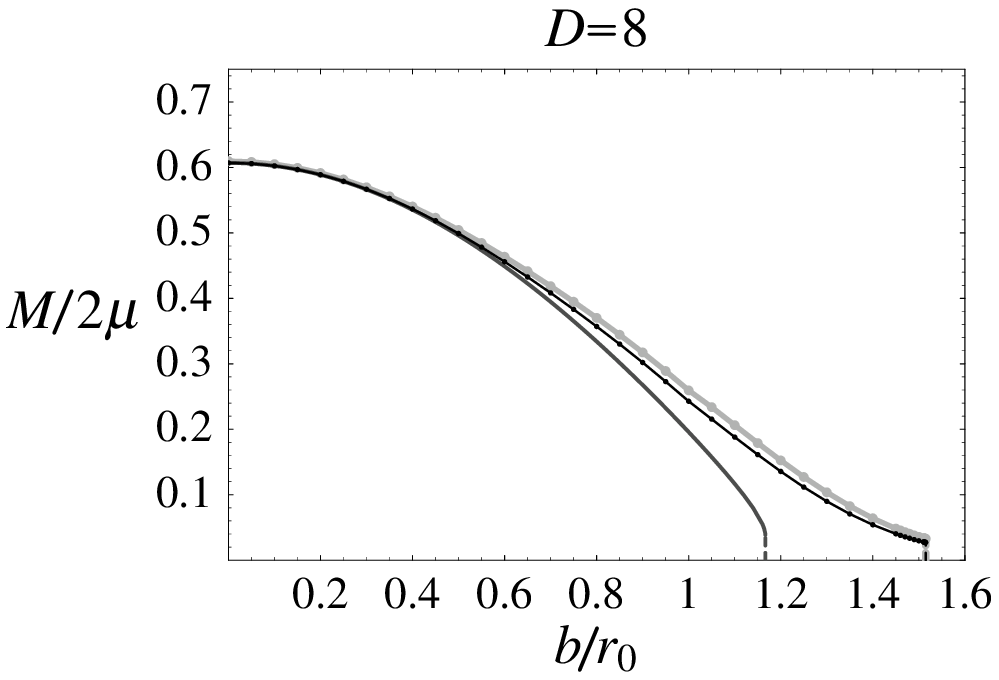}\hspace{10mm}
\includegraphics[width=0.3\textwidth]{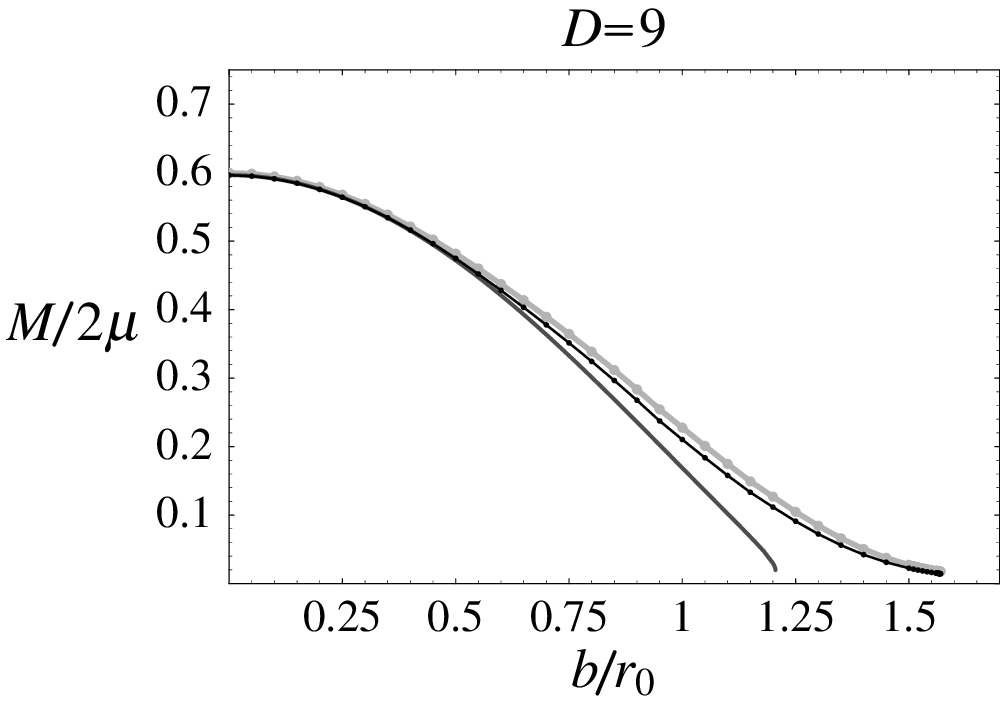}
\\
\vspace{3mm}
\includegraphics[width=0.3\textwidth]{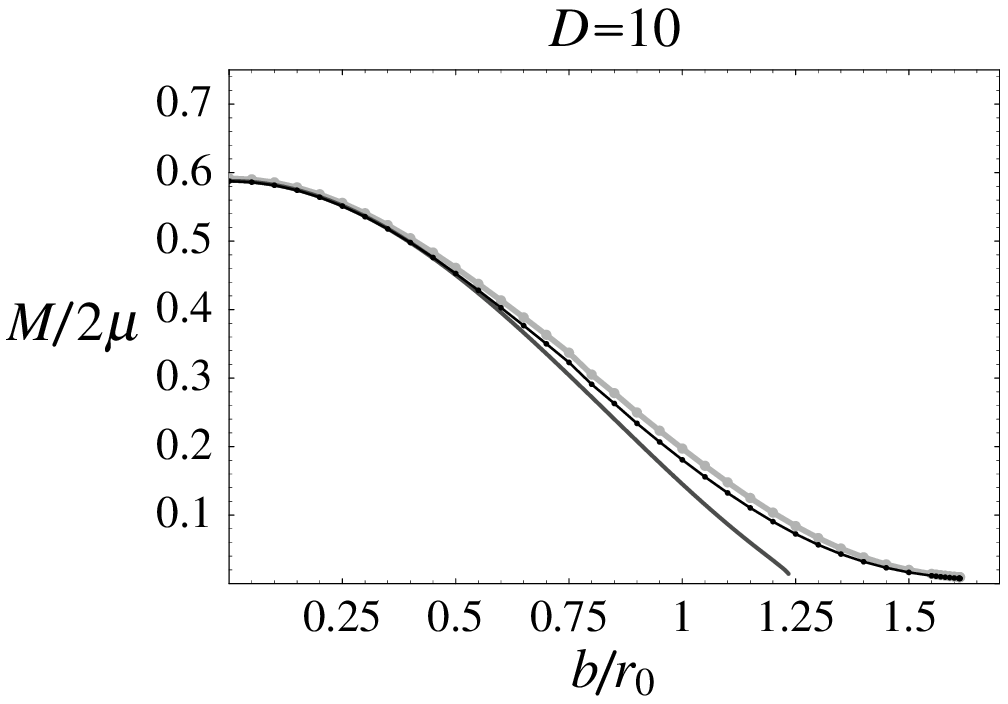}\hspace{10mm}
\includegraphics[width=0.3\textwidth]{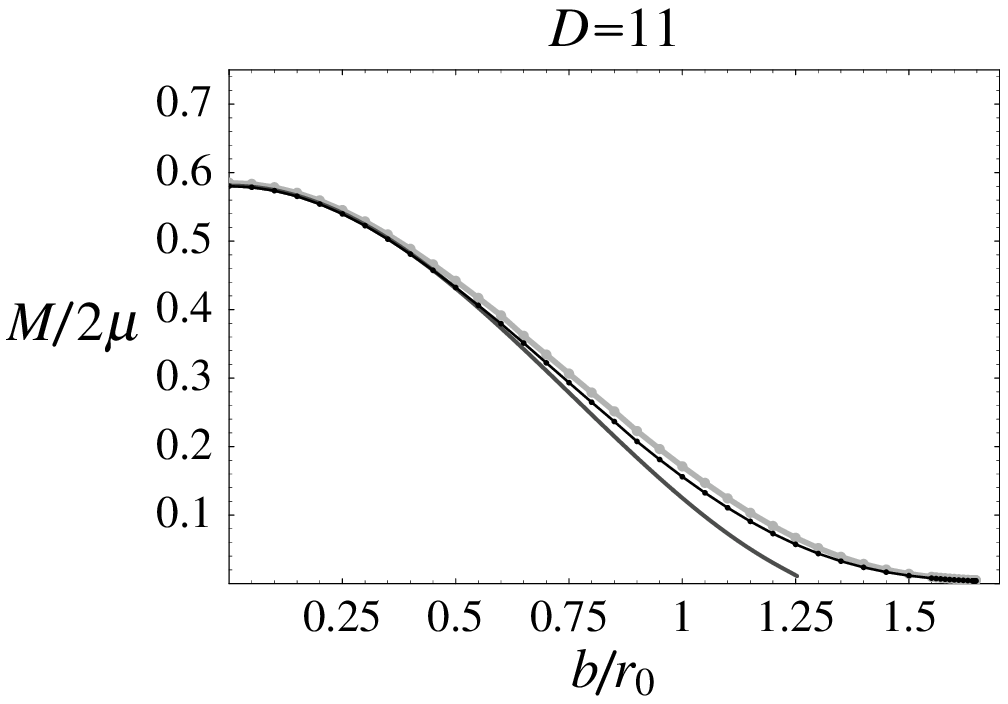}
}
\caption{The rigorous lower bound $M_{lb}$ of the final irreducible mass (black lines)
and the indicator of the trapped energy $M_{\rm AH}$ (light gray lines) for $D=4,...,11$.
The previous value of the AH mass $\hat{M}_{\rm AH}$ of~\cite{EG02, YN03}
are also shown by dark gray lines.   }
\label{Mlb_MAH}
\end{figure}

This $M_{lb}$ as a function of $b$ is shown in Fig.~\ref{Mlb_MAH}.
The non-rigorous bound $M_{\rm AH}$ is also included, because
under some additional assumptions it may still provide the energy
trapped by the produced black hole. For example, the Hawking
quasi-local mass~\cite{H68} calculated on the AH coincides with
$M_{\rm AH}$. The old-slice value of the AH mass $\hat{M}_{\rm
AH}$ as found in \cite{EG02,YN03} is also shown. We see that
$M_{lb}$ and $M_{\rm AH}$ take close values, $M_{\rm AH}$ being
slightly larger. Although $M_{lb}$ and $M_{\rm AH}$ are close to
$\hat{M}_{\rm AH}$ for small $b$, they becomes significantly
larger around $b=\hat{b}_{max}$, especially in the
higher-dimensional cases.

Now we consider the mass $M$ and the angular momentum $J$ of the
final Kerr black hole which are allowed by the area theorem:
\begin{equation}
M_{irr}\ge M_{lb}. \label{condition}
\end{equation}
Here $M_{irr}=M_{irr}(M,J)$ is the irreducible mass of the Kerr
black hole, which is defined, just as in four dimensions, as the
mass of a Schwarzschild black hole having the same horizon area.
It is thus related to the Kerr black hole horizon area $A_{Kerr}$ by the
formula:
$$
A_{Kerr}=\Omega_{D-3}r_{h}^{D-3}(M_{irr}).
$$
The left-hand side of this equation can be easily computed using
the explicit $D$-dimensional Kerr black hole metric \cite{MP86},
which gives the relation
\begin{equation}
r_h^{D-2}(M_{irr})=r_h^{D-3}(M)r_k(M,J).
\label{irreducible}
\end{equation}
Here $r_k(M,J)$ is the Kerr black hole horizon radius which
satisfies the following equation~\cite{MP86}:
\begin{equation}
r_k^2(M,J)+\left[\frac{(D-2)J}{2M}\right]^2=r_h^{D-3}(M)r_k^{5-D}(M,J).
\label{Kerr_radius}
\end{equation}

The total energy and the angular momentum of the system before the
collision are $2\mu$ and $b\mu$, respectively. Denoting
\begin{align}
\xi&=M/2\mu,\\
\zeta&=J/b\mu,
\end{align}
the final state of the produced black hole will be specified by a
point in $(\xi,\zeta)$-plane with $0\le \xi \le 1$ and $0\le \zeta
\le 1$.

For $D=4,5$ there exists an upper limit on the black hole angular
momentum for a fixed mass \cite{MP86}:
\begin{equation}
J\le J_{\star}(M)\equiv\left\{
\begin{array}{ll}
(1/2)Mr_h(M)& (D=4),\\
(2/3)Mr_h(M)& (D=5),
\end{array}
\right.
\end{equation}
 Region (i) consisting of points satisfying the opposite condition $J>J_{\star}(M)$ should be
{\it a priori} excluded from the $(\xi,\zeta)$-diagram.

According to the above discussion, region (ii) corresponding to
black holes violating the area theorem (\ref{condition}) can also
be excluded. The remaining points constitute the allowed region
(iii).

\begin{figure}[t]
\centering {
\includegraphics[width=0.3\textwidth]{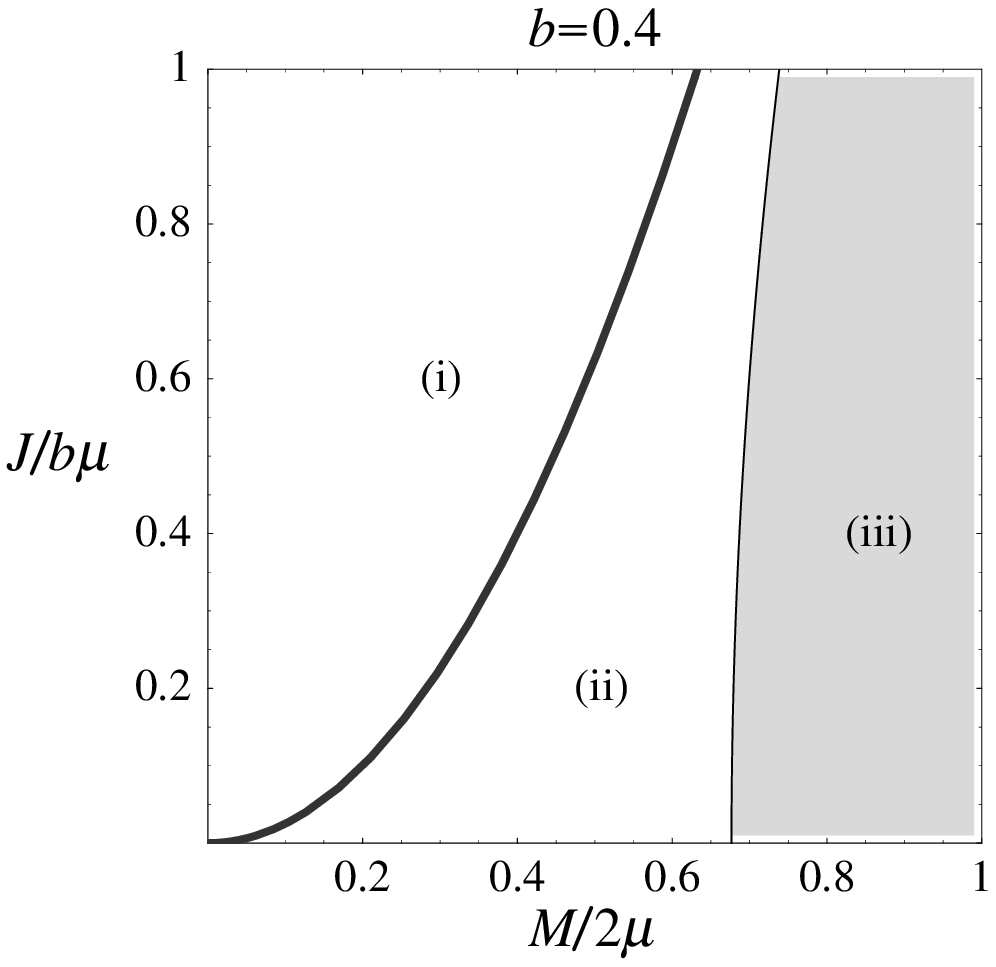}
\hspace{3mm}
\includegraphics[width=0.3\textwidth]{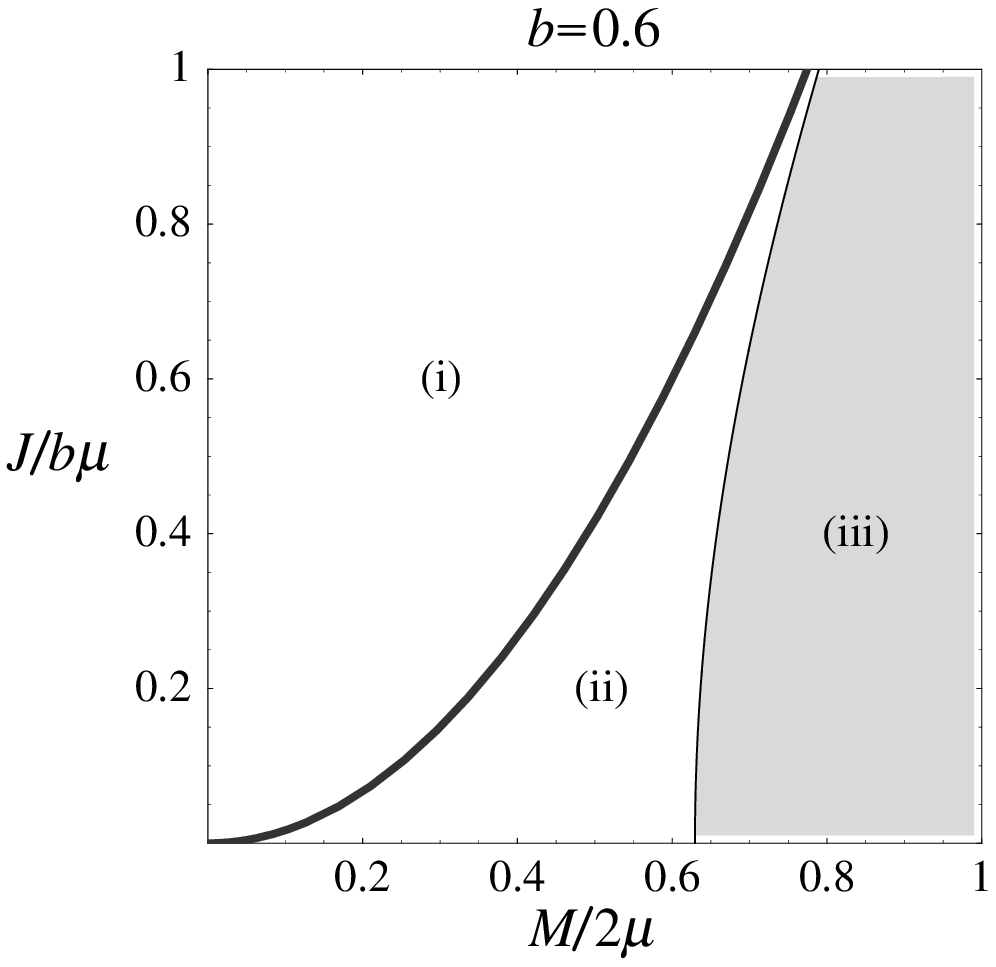}
\hspace{3mm}
\includegraphics[width=0.3\textwidth]{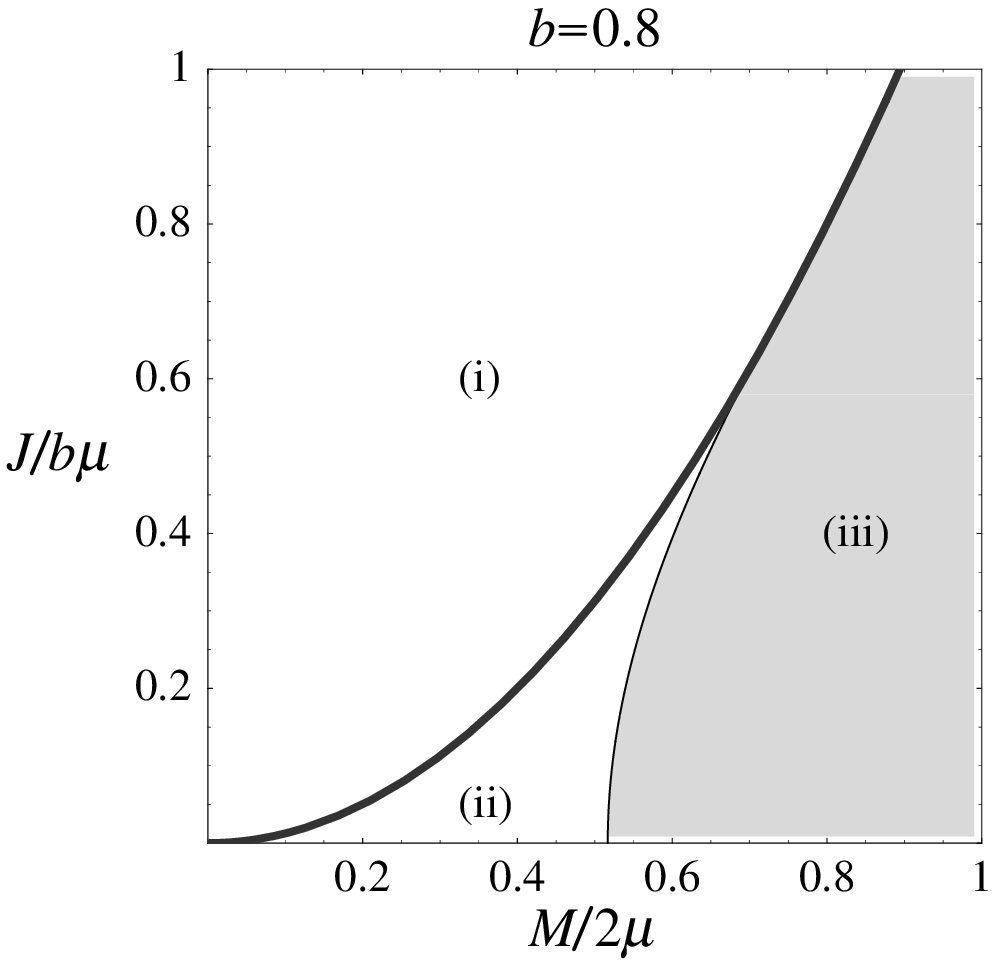}
} \caption{The regions (i), (ii) and (iii) in the
$(\xi,\zeta)$-plane for $b=0.4,0.6,0.8$ in the $D=4$ case. }
\label{MJ_allowed_D4}
\end{figure}
\begin{figure}[t]
\centering {
\includegraphics[width=0.3\textwidth]{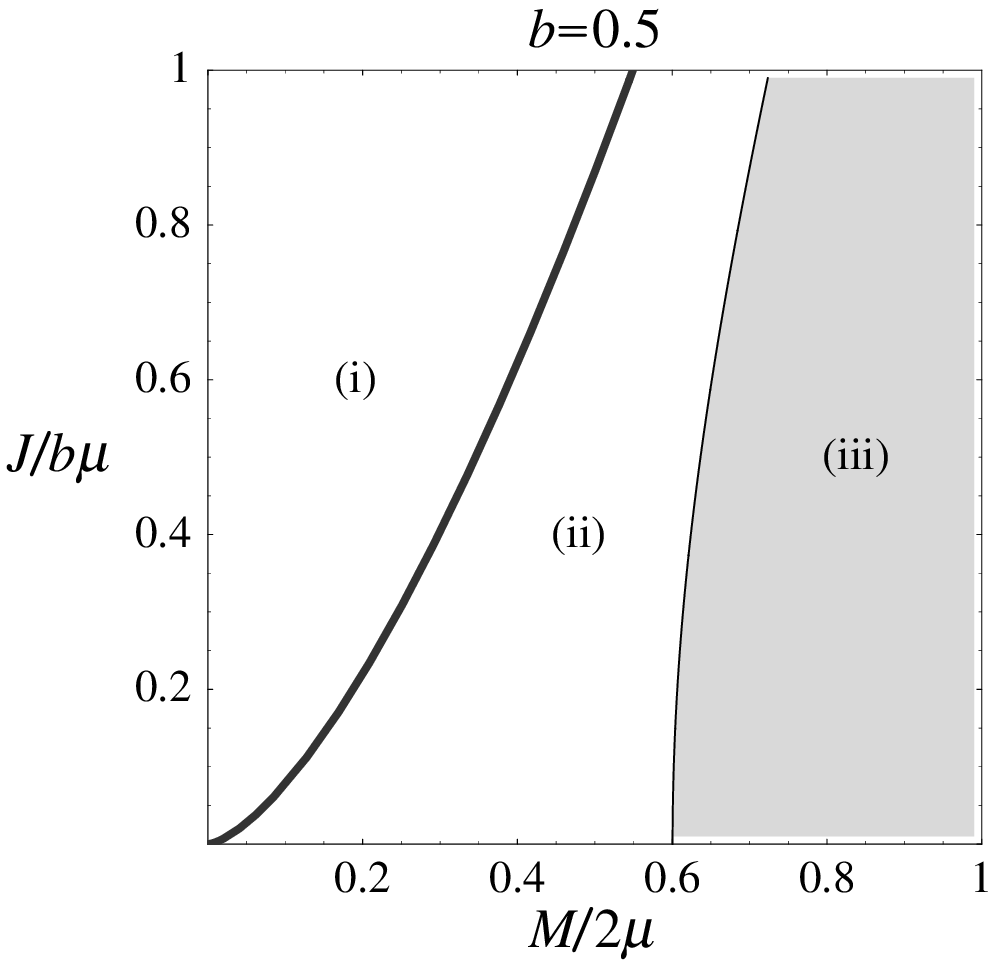}
\hspace{3mm}
\includegraphics[width=0.3\textwidth]{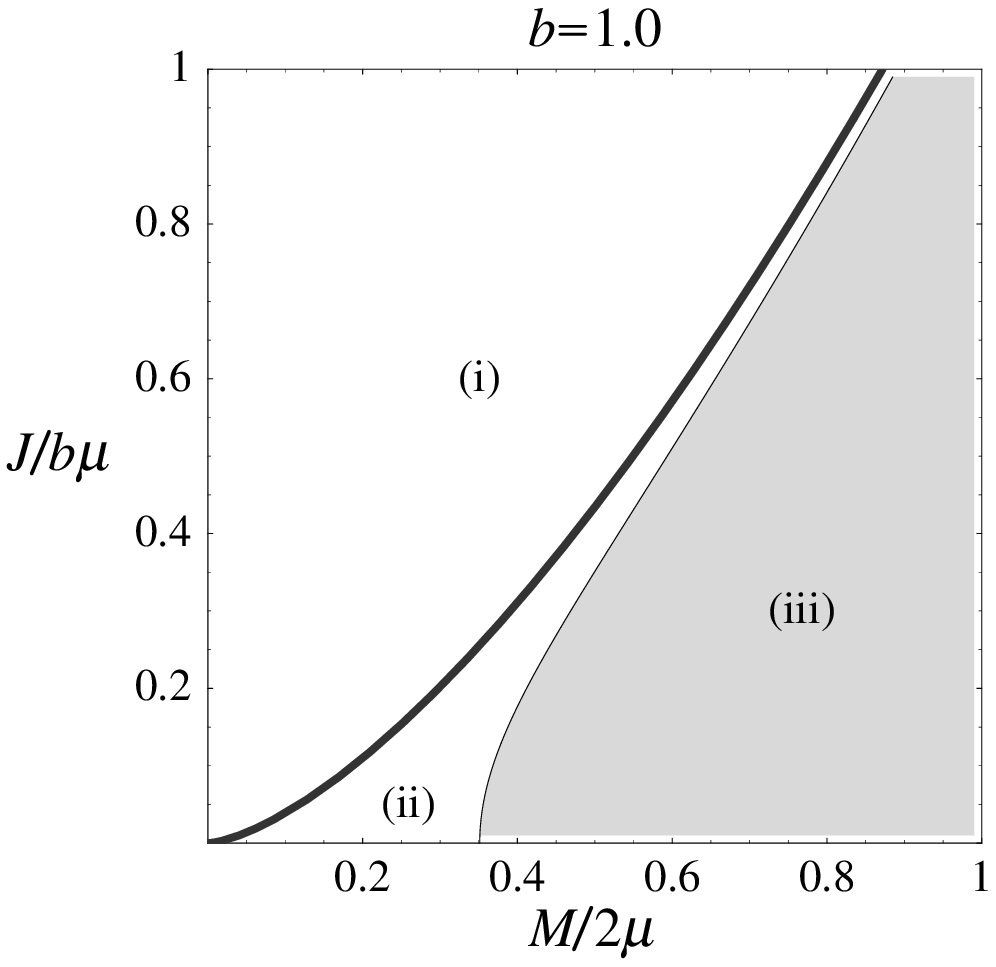}
\hspace{3mm}
\includegraphics[width=0.3\textwidth]{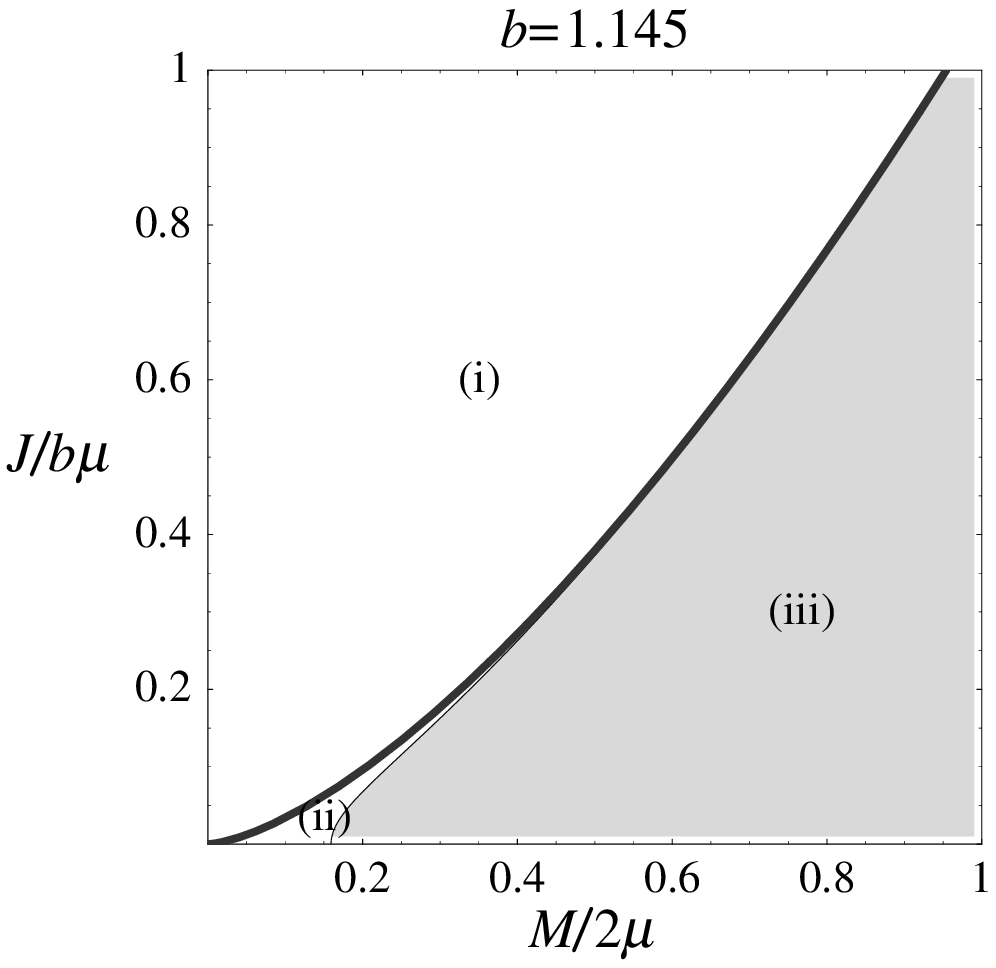}
} \caption{The regions (i), (ii) and (iii) in the
$(\xi,\zeta)$-plane for $b=0.5,1.0,1.145$ in the $D=5$ case. }
\label{MJ_allowed_D5}
\end{figure}

\begin{figure}[t]
\centering {
\includegraphics[width=0.3\textwidth]{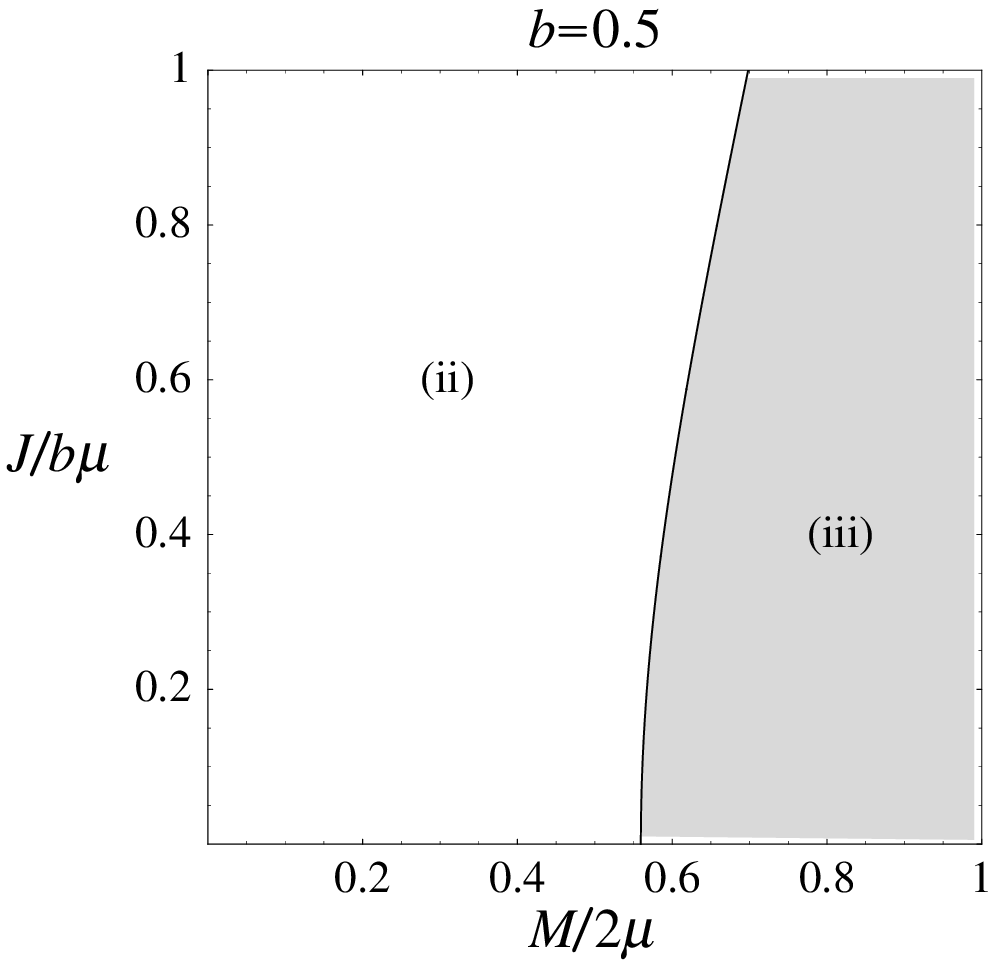}
\hspace{3mm}
\includegraphics[width=0.3\textwidth]{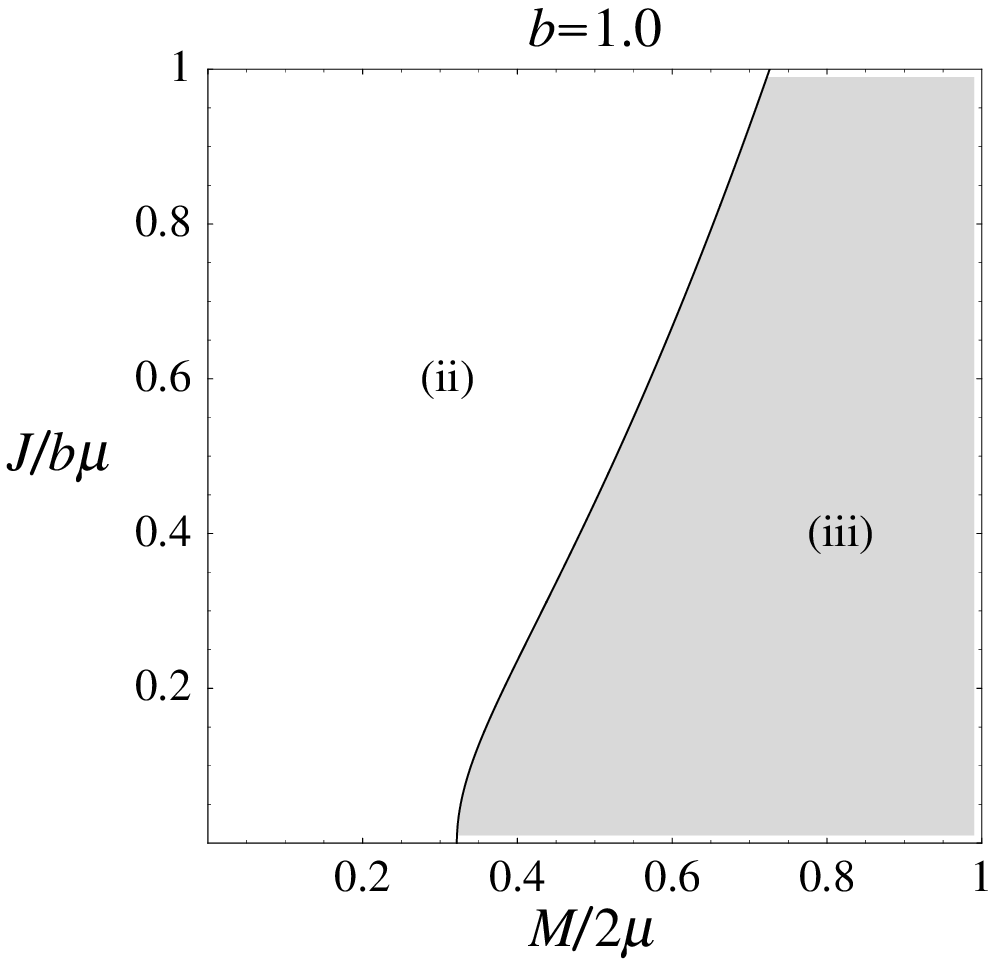}
\hspace{3mm}
\includegraphics[width=0.3\textwidth]{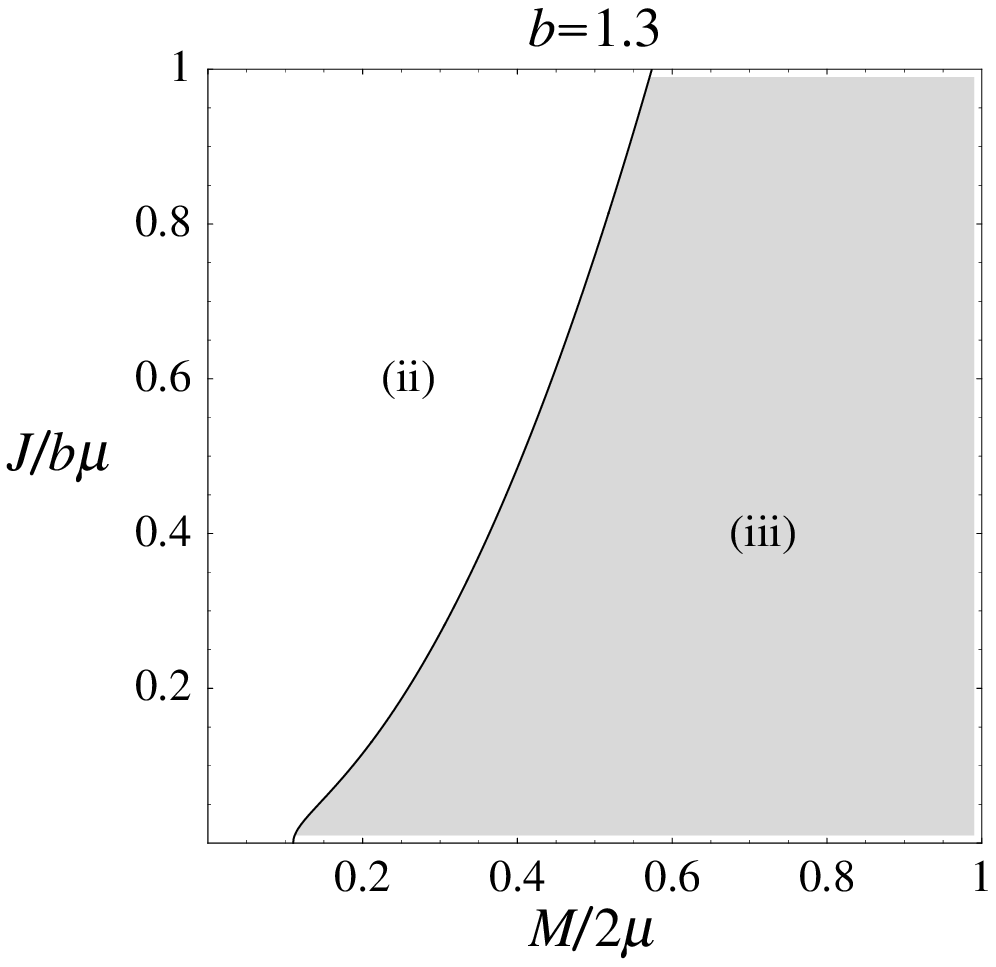}
} \caption{The regions (ii) and (iii) in the $(\xi,\zeta)$-plane
for $b=0.5,1.0,1.3$ in the $D=6$ case.   } \label{MJ_allowed_D6}
\end{figure}

\begin{figure}[t]
\centering {
\includegraphics[width=0.3\textwidth]{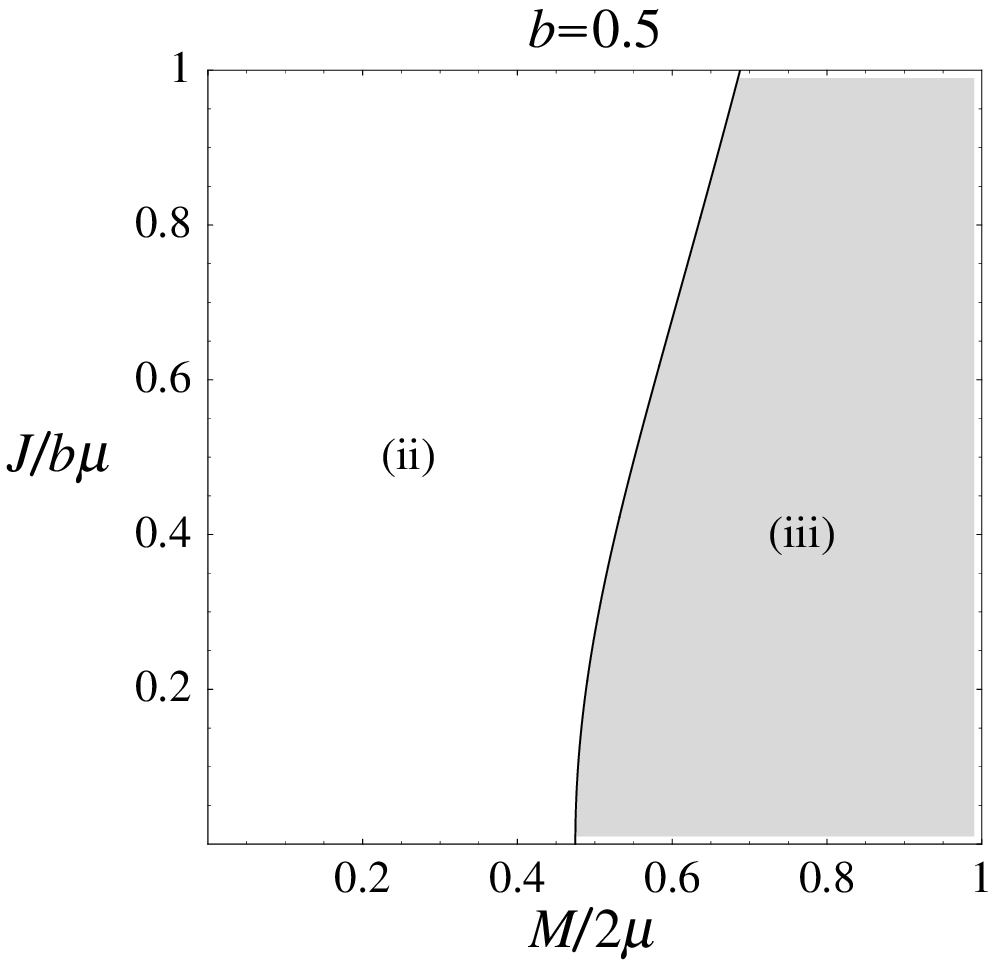}
\hspace{3mm}
\includegraphics[width=0.3\textwidth]{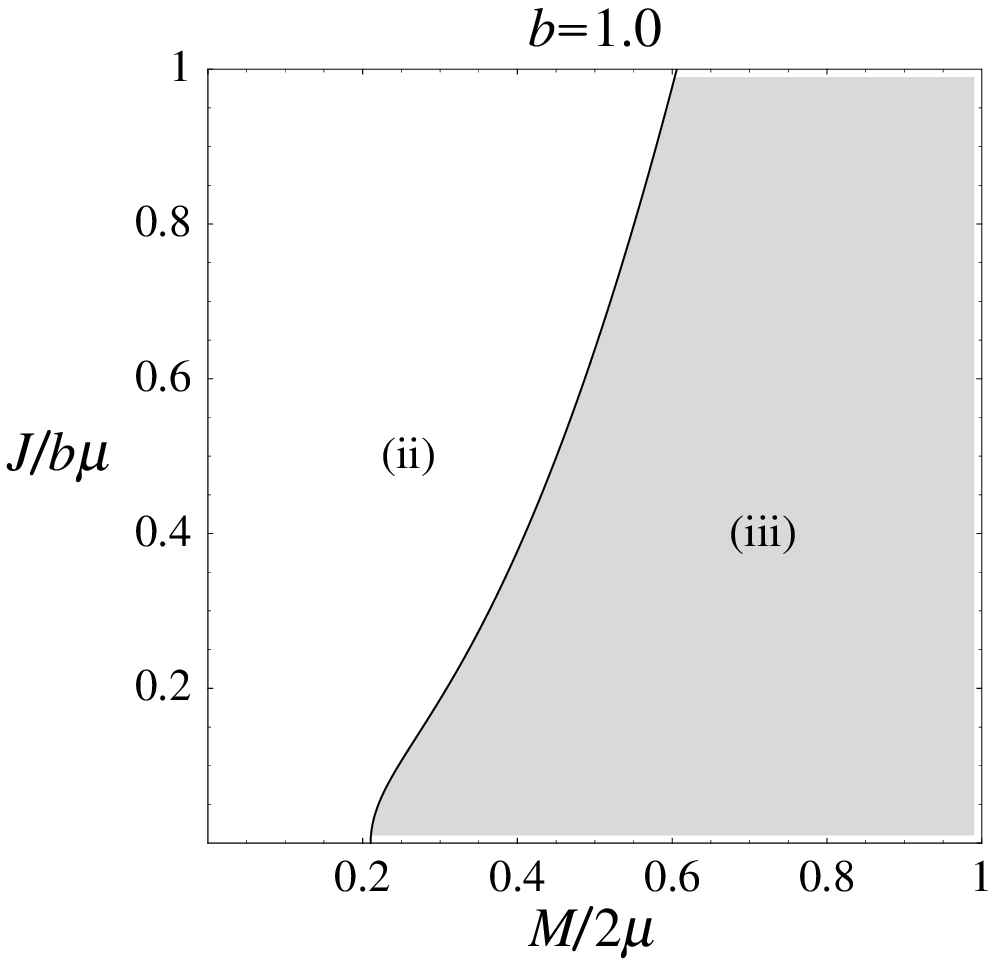}
\hspace{3mm}
\includegraphics[width=0.3\textwidth]{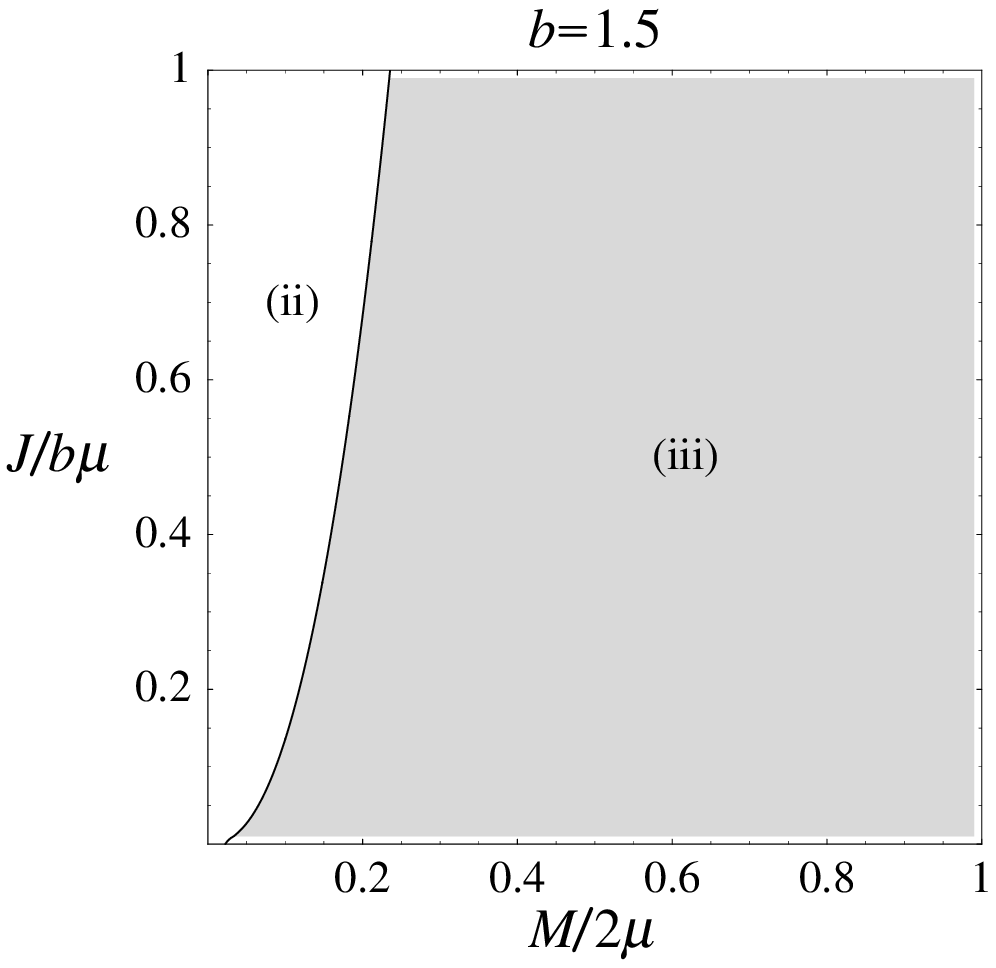}
} \caption{The regions (ii) and (iii) in the $(\xi,\zeta)$-plane
for $b=0.5,1.0,1.5$ in the $D=9$ case. } \label{MJ_allowed_D9}
\end{figure}

Figures \ref{MJ_allowed_D4}-\ref{MJ_allowed_D9}
show regions (i), (ii), (iii) for $D=4,5,6,$ and $9$ for some
selected values of $b$. We see that the condition $M_{irr}>M_{lb}$
gives a stronger restriction on the final state $(\xi,\zeta)$ than
the simple condition $M>M_{lb}$. This difference becomes quite
noticeable especially for $b\simeq b_{max}$ in the $D=4$ and $5$
cases. This result indicates that $M$ should be quite a bit larger
than $M_{lb}$ at $b\simeq b_{max}$, because almost 100\% angular
momentum should be radiated away if $M\simeq M_{lb}$, which would
be quite unnatural.

Unfortunately, we cannot find a non-trivial upper bound for the
angular momentum $J$ of the final Kerr black hole. (If the
boundary of region (iii) intersected the $\xi=1$ line at
$\zeta<1$, we would be able to find such a bound.) On the other
hand, it is interesting to note that our results are quite
consistent with previous numerical simulations of gravitational
collapse of rapidly rotating bodies in four dimensions (see
\cite{SS04} and references therein). In these works, the authors
found a necessary condition for black hole formation expressed as
\begin{equation}
q\equiv J_{\rm system}/J_{\star}(M_{\rm system})\lesssim 1,
\label{q_criterion}
\end{equation}
where $M_{\rm system}$ and $J_{\rm system}$ are the total
gravitational mass and angular momentum of the system. In our
system, the value of $q$ at $b=b_{max}$ is
\begin{equation}
q=\left\{
\begin{array}{ll}
 0.84 &  (D=4),\\
0.93  &  (D=5),
\end{array}
\right. \label{q_values}
\end{equation}
which is in agreement with~\eqref{q_criterion}. It should be
pointed out that for the five-dimensional black ring
solutions~\cite{ER02} there is no upper bound on $q$, and thus we
expect that criterion~\eqref{q_criterion} in the five-dimensional
case holds only for formation of the AH with spherical topology.

\section{Summary and discussion}

In this paper, we have analyzed the AH formation in the
high-energy particle collision using a new slice $u=0, v>0$ and
$v=0,u>0$, which lies to the future of the slice $u=0,v<0$ and
$v=0, u<0$ used in the previous studies of \cite{EG02, YN03}. Our
main results are summarized in Table~II. Compared to the previous
results for $\hat{b}_{max}$, we have obtained maximal impact
parameters $b_{max}$ of the AH formation larger by 18-30\% in the
higher-dimensional cases. These results lead to 40-70\% larger
cross section of the AH formation, the present value being
$\sigma_{\rm AH}\simeq 3\pi \left[r_h(2\mu)\right]^2$ for large
$D$.

We have also estimated the mass $M$ and angular momentum $J$ of
the final state of the produced black hole, as allowed by the area
theorem $M_{irr}>M_{lb}$. This condition provides a stricter
restriction on the final $M$ and $J$ than the simple condition
$M>M_{lb}$, and becomes especially effective for large $b$ in the
$D=4$ and $5$ cases, when our results indicate that the final mass $M$
should be significantly larger than $M_{lb}$. We also found that
Eq.~\eqref{q_values} gives a necessary condition for the AH
formation in the $D=4$ and $5$ cases, which is consistent with the
various numerical simulations of the gravitational collapse.

Our analysis provides the most precise data on the cross section
of the black hole production in high-energy particle collisions to date.
Using our new results, various phenomenological discussions that used the
results of \cite{YN03} (such as e.g.~\cite{improve}) or relied on
the more rough estimate (\ref{rough}) (e.g.~\cite{naive} and many
others) can be improved. The present investigation is a necessary
step towards the final understanding of the semi-classical signals
that would be observed in future-planned accelerators.

It should be stressed that the estimates on $M$ and $M_{irr}$
provided by our analysis give only rigorous upper bounds on the
amount of emitted gravitational radiation. The real amount is
likely to be smaller than suggested by these estimates, by a
factor of a few. The work of D'Eath and Payne~\cite{DP92} gives an
idea about the size of this effect. In their analysis of
axisymmetric collision of two Aichelburg-Sexl particles, they
calculated the evolution of the gravitational field far away from
the center using $\gamma^{-1}$ as a small parameter, and derived
the news function near the symmetry axis to the second order.
Assuming the azimuthal pattern of the gravitational radiation,
they estimated the energy loss to be 16\%, which should be
compared to the rigorous upper bound of 29\% provided by the AH
method\footnote{It should be also noted that D'Eath and Payne's
estimate does not take into account additional gravitational
radiation from the center of the system, which cannot be evaluated
by this method.}. It is natural to expect that reduction of
comparable size will occur in all dimensions. It should be
mentioned, however, that a recent calculation~\cite{CDL03} based
on an ``instantaneous collision" approximation in linearized
gravity predicts that gravitational wave emission becomes highly
suppressed in higher dimensions (up to 0.001\% in $D=10$), which
in our opinion is unlikely. Still another setup
\cite{Schwarzschild} models the collision by a lightlike particle
falling into a Schwarzschild black hole and gives estimates which
are closer to our values (8\% in $D=10$). We point out that all
these works have problems such as ignoring the nonlinearity of the
system, or the setup is too far from the realistic one. Analysis
without approximations remains an important open problem.

The ultimate goal of such analyses, which is left for the future,
is to determine the spacetime structure after the collision,
i.e., in region IV ($u>0, v>0$). This would clarify the precise
 maximal impact parameter of the black hole formation and the
relation between the values of $(M,J)$ of the final state and the
impact parameter $b$. If this is completed, we will be able to
obtain quite accurate semi-classical predictions by using the
existing studies of the greybody factors.

\acknowledgments
H.Y. acknowledges helpful comments of Seong Chan Park and Ken-ichi Nakao.
V.R. would like to thank Mihalis Dafermos and
Luciano Rezzolla for useful discussions. H.Y. thanks Nagoya University
21st Century COE Program (ORIUM) for financial support. The work of
V.R. was supported by Stichting FOM.

\end{document}